\documentclass[twocolumn,aps,showpacs,showkeys,prb,superscriptaddress,reprint]{revtex4-1}
\usepackage{natbib}
\usepackage{bm}
\usepackage{bbm}
\usepackage{textcase}
\usepackage{color}
\usepackage{graphicx}
\usepackage{amsmath}
\usepackage{amssymb}
\usepackage{url}
\usepackage[colorlinks=true,citecolor=blue,hyperindex,breaklinks]{hyperref} 

\hypersetup{pdftitle={Adaptive cluster approximation for reduced density-matrix functional theory}}
\hypersetup{pdfauthor={Robert Schade, Peter Bloechl}}
\hypersetup{pdfdisplaydoctitle}

\newcommand{\mat}[1]{{\boldsymbol #1}}

\begin{document}

\title{Adaptive cluster approximation for reduced density-matrix functional theory}

\author{Robert Schade}
\email{robert.schade@tu-clausthal.de}

\affiliation{Institute for Theoretical Physics, Clausthal University
  of Technology, Leibnizstr. 10, 38678 Clausthal-Zellerfeld, Germany}

\author{Peter E. Bl\"ochl}
\affiliation{Institute for Theoretical Physics, Clausthal University
  of Technology, Leibnizstr. 10, 38678 Clausthal-Zellerfeld, Germany}

\affiliation{Institute for Materials Physics,
  Georg-August-Universit\"at G\"ottingen, Friedrich-Hund-Platz 1,
  37077 G\"ottingen, Germany} 
\date{\today}

\begin{abstract}
A method, called the adaptive cluster approximation (ACA), for single-impurity Anderson models is proposed. It is based on reduced density-matrix functional theory, where the one-particle reduced density matrix is used as the basic variable. The adaptive cluster approximation introduces a unitary transformation of the bath states such that the effect of the bath is concentrated to a small cluster around the impurity. For this small effective system one can then either calculate the reduced density-matrix functional numerically exact from Levy's constrained-search formalism or approximate it by an implicit approximation of the reduced density-matrix functional. The method is evaluated for single-impurity Anderson models with finite baths. The method converges rapidly to the exact result with the size of the effective bath.
\end{abstract}

\maketitle

\section{\label{sec:introduction}Introduction}
Reduced density-matrix functional theory\cite{gilbert75_prb12_2111,levy79_pnas76_6062,lieb83_ijqc24_243} (rDMFT) emerged recently as viable option to describe materials with strong electronic correlations. It can be seen as a relative to density-functional theory\cite{hohenberg64_pr136_B864,kohn65_pr140_1133} (DFT) that treats the one-particle reduced density matrix as the basic quantity instead of the electron density. In this sense rDMFT emphasizes orbital occupations that are more natural for the description of correlated materials. 

Like the exchange-correlation functional of DFT, the effort to evaluate the exact reduced density-matrix functional is  prohibitively high. 
Different strategies have been used to cope with this problem: 
Analogous to DFT, the many-particle problem can be encoded in approximate, parameterized density-matrix functionals that can be evaluated with a small computational effort. 
Parametrized functionals have been applied to models\cite{PhysRevB.61.1764,PhysRevB.63.115116,PhysRevB.66.155118,PhysRevB.67.035115,PhysRevB.69.085101,PhysRevA.79.040501,toews11_prb83_235101,benavidesriveros12_epjd66_274,toews12_prb86_245123,PhysRevB.94.045102,disabatino15_jcp143_24108,kamil16_prb93_085141} and real systems\cite{goedecker98_prl81_866,staroverov02_jcp117_2489,lathiotakis08_jcp128_184103,lathiotakis07_prb75_195120,gritsenko05_jcp122_204102,marques08_pra77_32509,sharma13_prl110_116403,shinohara1,shinohara2,PhysRevB.94.155141,QUA:QUA10707,QUA:QUA20858,QUA:QUA24663,doi:10.1021/ct300414t,Pernal2016}.  

Levy's constrained-search algorithm\cite{levy79_pnas76_6062,bloechl11_prb84_205101} describes a constrained optimization problem in the space of many-particle wave functions. This method inherits the difficulties of the many-particle problem. While being exact in practice it is restricted to rather small system sizes. 
Functionals that require the solution of an internal optimization problem like Levy's constrained-search algorithm or other implicit density-matrix functionals suffer from a unfavorable scaling of the computational complexity with the total system size.

The electron-electron interaction in a single-impurity Anderson model\cite{anderson61_pr124_41} (SIAM)  is limited to only few one-particle basis states but the entire system is a many-particle quantum problem. These interacting states are named the impurity. The remaining non-interacting states are called the bath. Therefore, methods that allow to create a smaller effective bath are highly desirable. 
An example for such an approach is the two-level approximation\cite{toews11_prb83_235101,toews12_prb86_245123} by T\"ows et al. The main idea of the two-level-approximation is to introduce a unitary transformation of the bath states, such that only two of the transformed bath states have finite density-matrix elements with the impurity. All other basis states are neglected, which gives an effective system of four basis states, two impurity states and two basis states, for which the density-matrix functional is known. Consequently the two-level-approximation is limited to impurities with two spin-orbitals and an effective bath consisting of two spin-orbitals.

In this paper we introduce a method, named adaptive cluster approximation (ACA), that can handle impurity problems within rDMFT with an arbitrary number of effective bath states/levels, arbitrary impurity sizes and multi-band interactions. The method sets up a unitary transformation between the non-interacting basis states, that aims at minimizing the subsequent truncation error of bath states. This creates a smaller effective cluster composed of the impurity and an effective bath for which one has to evaluate the density-matrix functional.

The paper is organized as follows. In section~\ref{sec:rDMF}, we first describe the basic ideas of rDMFT. Then we present the adaptive cluster approximation in sec.~\ref{sec:adaptive_cluster_approximations}. We describe relations to existing methods and present exact limits. In section~\ref{sec:numerical_methods} the numerical methodology is presented. In section~\ref{sec:siam} we describe applications of the method to single-impurity Anderson models with a finite bath and compare to numerically exact ground state from exact diagonalization. We explore the dependence of results of the ACA on different bath truncations.  In section~\ref{sec:siam_beyond_all} we investigate under which conditions the ACA is exact for single-impurity Anderson models. We explore SIAMs with very large baths that approach the limit of a continuous density of states and with multi-orbital impurities.

\section{\label{sec:rDMF}Reduced density-matrix functional theory}
A many-particle Hamiltonian $\hat H$ can be written as a sum of a non-interacting part $\hat h$ and the general two-particle interaction Hamiltonian $\hat W$,
\begin{eqnarray}
  \hat H&=&\hat h+\hat W.
\end{eqnarray}
In an orthonormal one-particle basis set, the non-interacting part $\hat h$ of the Hamiltonian can be written as
\begin{eqnarray}
  \hat h&=&\sum_{a,b}h_{a,b} \hat c^\dagger_a \hat c_b
\end{eqnarray}
and the general two-particle interaction Hamiltonian $\hat W$ as
\begin{eqnarray}
  \hat W&=&\frac{1}{2}\sum_{a,b,c,d}U_{a,b,d,c} \hat c^\dagger_a \hat c^\dagger_b\hat c_c \hat c_d
\end{eqnarray}
using the creation and annihilation operators $\hat c^\dagger_a$ and $\hat c_a$.
The one-particle reduced density matrix $\rho_{b,a}$ of an ensemble of normalized fermionic many-particle wave functions $|\Psi_i\rangle$ with probabilities $P_i$ ($0\leq P_i$, $\sum_i P_i=1$) is defined as
\begin{eqnarray}
  \rho_{b,a}&=& \sum_i P_i \langle \Psi_i| \hat c^\dagger_a \hat c_b|\Psi_i\rangle.\label{eq:denmat_from_ensemble}
\end{eqnarray}
All hermitian matrices $\mat{\rho}$ that can be generated by Eq.~\eqref{eq:denmat_from_ensemble}  from an ensemble of normalized fermionic many-particle wave functions $|\Psi_i\rangle$ with probabilities $P_i$ are called ensemble representable. 
Coleman\cite{coleman63_rmp35_668} has shown that ensemble representability is equivalent to the property that the eigenvalues of the one-particle reduced density matrix, called occupations by L\"owdin\cite{loewdin55_pr97_1474}, are between zero and one. In other words $\mat{\rho}$ and $\mat{1}-\mat{\rho}$ are positive semi-definite, in short $\mat{0}\leq \mat{\rho} \leq \mat{1}$.

Within reduced density-matrix functional theory\cite{gilbert75_prb12_2111,levy79_pnas76_6062,lieb83_ijqc24_243} (rDMFT) the ground-state energy of a N-particle system can be written as 
\begin{equation}
  \label{eq:omega_min}
  E_{N}(\hat h+\hat W)=\min_{\mat{\rho},\mat{0}\leq \mat{\rho} \leq \mat{1},\mathrm{Tr}(\mat{\rho})=N}\left\{ \mathrm{Tr}[\mat{\rho} \mat{h}]+F^{\hat W}[\mat{\rho}]\right\},
\end{equation}
where the minimization is performed over all ensemble-representable one-particle reduced density matrices $\mat{\rho}$ with $N$ particles. The dependence of the interaction energy on the one-particle density matrix is encoded in the density-matrix functional $F^{\hat W}_{}[\mat{\rho}]$ in Eq.~\eqref{eq:omega_min}.
This functional is a universal functional of the one-particle reduced density matrix in the sense that it does not depend on the external one-particle potential of the system\cite{gilbert75_prb12_2111}. Levy\cite{levy79_pnas76_6062} and Valone\cite{1.440249} have shown that the density-matrix functional can be obtained from a constrained minimization over an ensemble of orthonormal fermionic many-particle wave functions $|\Psi_i\rangle$ and ensemble probabilities $P_i$ with $0\leq P_i$ and $\sum_i P_i=1$ as 
\begin{equation}
  \label{eq:rDMF}
  F^{\hat W}_{}[\mat{\rho}]=\min_{\{P_i,|\Psi_i\rangle\} \rightarrow \mat{\rho}} \sum_i P_i \langle \Psi_i| \hat W|\Psi_i\rangle .
\end{equation}
With $\{P_i,|\Psi_i\rangle\} \rightarrow \mat{\rho}$ we denote  the set of ensembles with a given one-particle reduced density matrix $\mat{\rho}$ according to Eq.~\eqref{eq:denmat_from_ensemble}.

For one-particle reduced density matrices that correspond to non-degenerate ground states, the density-matrix functional can be written as minimization over just one many-particle wave function, that is $P_1=1$ and $P_i=0$ for $i>1$, in the form
\begin{equation}
  \label{eq:rDMF2}
  F^{\hat W}_{\mathrm{single}\ |\Psi\rangle}[\mat{\rho}]=\min_{|\Psi\rangle \rightarrow \mat{\rho}} \langle \Psi| \hat W|\Psi\rangle.
\end{equation}
For all systems, the relation
\begin{equation}
  \label{eq:rdmfcomp}
  F^{\hat W}_{}[\mat{\rho}] \leq F^{\hat W}_{\mathrm{single}\ |\Psi\rangle}[\mat{\rho}]
\end{equation}
holds. It should also be noted here, that rDMFT, like DFT, is a ground state-theory. The generalization of rDMFT to finite temperatures is straight-forward\cite{baldsiefen15_pra92_052514}.
There is no direct way to extract dynamical spectral functions but additional approximations need to be introduced and the physical content of those is still under intense discussion\cite{sharma13_prl110_116403,disabatino15_jcp143_24108,PhysRevB.94.155141}. Thus we report here only on ground-state properties such as energies and orbital occupations.

\section{\label{sec:adaptive_cluster_approximations}Adaptive cluster approximation}
\subsection{\label{sec:adaptive_cluster_approximations_trans}Transformation of the bath basis states}
The solution of the rDMFT-minimization problem given by Eq.~\eqref{eq:omega_min} requires the calculation of the reduced density-matrix functional $F^{\hat W}[\mat{\rho}]$ for a given one-particle reduced density matrix $\mat{\rho}$ in every minimization step. 
Unfortunately, calculating the density-matrix functional by Eq.~\eqref{eq:rDMF} scales exponentially with the size $N_\chi$ of the one-particle basis, even for a local interaction. This is the case because the one-particle reduced density matrix $\mat{\rho}$ includes all one-particle basis states of the system.
If only a subset of the one-particle states take part in the two-particle interaction as for a single-impurity Anderson model, we can subdivide one-particle states in two disjoint sets: a set $C_\mathrm{imp}$ of interacting orbitals (impurity) and a set $C_\mathrm{bath}$ of non-interacting orbitals (bath). The interaction then has the form:
\begin{eqnarray}
  \hat W&=&\frac{1}{2}\sum_{a,b,c,d\in C_\mathrm{imp}}U_{a,b,d,c} \hat c^\dagger_a \hat c^\dagger_b\hat c_c \hat c_d.
\end{eqnarray}
In order to set the stage, let us consider the limiting case of a density-matrix functional $F^{\hat W}[\mat{\rho}]$ with a density matrix that is block-diagonal with respect to $C_\mathrm{imp}$ and $C_\mathrm{bath}$, i.e.
\begin{eqnarray}
  \label{eq:block_rho0}
  \mat{\rho}&=&\begin{pmatrix}\mat{\rho}_{\mathrm{imp,imp}} & \mat{0} \\ \mat{0} & \mat{\rho}_{\mathrm{bath,bath}} \end{pmatrix}.
\end{eqnarray}
The density-matrix functional of this density matrix is independent of the block $\mat{\rho}_{\mathrm{bath,bath}}$ and can be calculated considering $\mat{\rho}_{\mathrm{imp,imp}}$ alone,
\begin{eqnarray}
  F^{\hat W}[\mat{\rho}]&=&F^{\hat W}[\mat{\rho}_{\mathrm{imp,imp}}].
\end{eqnarray}
The proof of this statement is provided in appendix~\ref{appendix:proof_truncation}.
In the case of a general one-particle reduced density matrix
\begin{eqnarray}
  \label{eq:rho_as_blocks}
  \mat{\rho}&=&\begin{pmatrix}\mat{\rho}_{\mathrm{imp},\mathrm{imp}} & \mat{\rho}_{\mathrm{imp},\mathrm{bath}} \\ \mat{\rho}_{\mathrm{imp},\mathrm{bath}}^\dagger & \mat{\rho}_{\mathrm{bath},\mathrm{bath}} \end{pmatrix},
\end{eqnarray}
we additionally rely on the invariance of the density-matrix functional with transformations:
\begin{eqnarray}
  \label{eq:unitary_trans_exact_rdmf}
  F^{\hat W}[\mat{\rho}]&=&F^{\hat{W}}[\mat{U}^\dagger \mat{\rho} \mat{U}],
\end{eqnarray}
where $\mat{U}$ is a unitary $N_\chi \times N_\chi$ matrix of the form 
\begin{equation}
  \mat{U}= \begin{pmatrix}\mat{1}_{N_\mathrm{imp}} & \mat{0} \\ \mat{0} & \mat{U}^\dagger_\text{bath} \end{pmatrix}.
\end{equation}
Here $\mat{1}_{N_\mathrm{imp}}$ is a $N_\mathrm{imp} \times N_\mathrm{imp}$ unit matrix and $\mat{1}_{N_\mathrm{imp}}$ is a $N_\mathrm{imp} \times N_\mathrm{imp}$ unit matrix
$N_\mathrm{imp}$ is the number of interacting states making up the impurity. 
The proof of this statement is provided in the supplemental material at [URL will be inserted by publisher].
We exploit this freedom to transform the density matrix to a banded form

\begin{widetext}
\begin{eqnarray}
  \label{eq:Ublocks}
  \tilde {\mat{\rho}}&=&\mat{U}^\dagger \mat{\rho} \mat{U}= \begin{pmatrix}\mat{1}_{N_\mathrm{imp}} & \mat{0} \\ \mat{0} & \mat{U}^\dagger_\text{bath} \end{pmatrix} \mat{\rho} \begin{pmatrix}\mat{1}_{N_\mathrm{imp}} & \mat{0} \\ \mat{0} & \mat{U}_\mathrm{bath} \end{pmatrix}\\
  &=&\left(
    \begin{array}{ccccccc}
      \mat{\rho}_{\mathrm{imp},\mathrm{imp}} & \tilde {\mat{\rho}}_{\mathrm{imp},\mathrm{bath}_1} & \mat{0} & \dots &  \\ 
      \tilde {\mat{\rho}}_{\mathrm{imp},\mathrm{bath}_1}^\dagger & \tilde {\mat{\rho}}_{\mathrm{bath}_1,\mathrm{bath}_1} & \tilde {\mat{\rho}}_{\mathrm{bath}_1,\mathrm{bath}_2} & \mat{0} & \dots  \\ 
      \mat{0} & \tilde{\mat{\rho}}_{\mathrm{bath}_1,\mathrm{bath}_2}^\dagger&\tilde {\mat{\rho}}_{\mathrm{bath}_2,\mathrm{bath}_2} & \tilde{\mat{\rho}}_{\mathrm{bath}_2,\mathrm{bath}_3} & \mat{0} & \dots \\
      \vdots & \mat{0} & \tilde{\mat{\rho}}_{\mathrm{bath}_2,\mathrm{bath}_3}^\dagger&\tilde {\mat{\rho}}_{\mathrm{bath}_3,\mathrm{bath}_3} & \tilde {\mat{\rho}}_{\mathrm{bath}_3,\mathrm{bath}_4} & \ddots &  \\
             & \vdots & \mat{0} & \tilde{\mat{\rho}}_{\mathrm{bath}_3,\mathrm{bath}_4}^\dagger& \tilde {\mat{\rho}}_{\mathrm{bath}_4,\mathrm{bath}_4}& \ddots \\
             &  & \vdots & \ddots & \ddots & \ddots \\
    \end{array}
  \right).
\label{eq:multi_level_trafo}
\end{eqnarray}
\end{widetext}

We construct the unitary transformation  $\mat{U}_\mathrm{bath}$ of the bath states iteratively. The construction of one step is outlined in the following: We write the one-particle reduced density matrix in block-form as in Eq.~\eqref{eq:rho_as_blocks},
\begin{eqnarray}
  \label{eq:rho_as_blocks2}
  \mat{\rho}&=&\begin{pmatrix}\mat{\rho}_{\mathrm{imp,imp}} & \mat{\rho}_{{\mathrm{imp,bath}_1}} & \mat{\rho}_{{\mathrm{imp,rest}}} \\ \mat{\rho}_{{\mathrm{imp,bath}_1}}^\dagger & \mat{\rho}_{ {\mathrm{bath}_1,\mathrm{bath}_1}} & \mat{\rho}_{ {\mathrm{bath}_1,\mathrm{rest}}} \\ \mat{\rho}_{{\mathrm{imp,rest}}}^\dagger & \mat{\rho}_{ {\mathrm{bath}_1,\mathrm{rest}}}^\dagger & \mat{\rho}_{{\mathrm{rest},\mathrm{rest}}} \end{pmatrix}.
\end{eqnarray}
The block $\mathrm{bath}_1$ contains $N_{\mathrm{bath}_1}$ states and the block $\mathrm{rest}$ the remaining $N_{\mathrm{rest}}$ states. 
Transforming the one-particle basis with a block-diagonal unitary transformation defined in Eq.~\eqref{eq:Ublocks}
\begin{eqnarray}
  \mat{U}&=&\begin{pmatrix}\mat{1}_{N_\mathrm{imp}} & \mat{0} & \mat{0} \\ \mat{0} & \mat{U}_{\mathrm{bath}_1,\mathrm{bath}_1} & \mat{U}_{\mathrm{bath}_1,\mathrm{rest}} \\ \mat{0} & \mat{U}_{\mathrm{rest},\mathrm{bath}_1} & \mat{U}_{\mathrm{rest},\mathrm{rest}} \end{pmatrix}
\end{eqnarray}
gives the transformed density matrix
\begin{equation}
  \tilde{\mat{ \rho}}=\mat{U}^\dagger \mat{\rho} \mat{U}=\begin{pmatrix}\mat{\rho}_{\mathrm{imp,imp}} & \tilde{\mat{\rho}}_{{\mathrm{imp,bath}_1}} & \tilde{\mat{ \rho}}_{{\mathrm{imp},\mathrm{rest}}} \\ \tilde{\mat{ \rho}}_{{\mathrm{imp,bath}_1}}^\dagger & \tilde{\mat{ \rho}}_{{\mathrm{bath}_1,\mathrm{bath}_1}} & \tilde{\mat{ \rho}}_{{\mathrm{bath}_1,\mathrm{rest}}} \\ \tilde{\mat{ \rho}}_{{\mathrm{imp},\mathrm{rest}}}^\dagger & \tilde{\mat{ \rho}}_{{\mathrm{bath}_1,\mathrm{rest}}}^\dagger & \tilde{\mat{ \rho}}_{{\mathrm{rest},\mathrm{rest}}} \end{pmatrix}.
\end{equation}
Now we determine the unitary matrix such that the coupling density-matrix elements $\tilde{\mat{ \rho}}_{{\mathrm{imp},\mathrm{rest}}}$ of the transformed density matrix vanish, i.e. $\tilde{\mat{ \rho}}_{{\mathrm{imp},\mathrm{rest}}}=\mat{0}$. This implicitly defines the unitary matrix $\mat{U}_{{\mathrm{bath}_1,\mathrm{rest}}}$ via 
\begin{equation}
  \label{eq:appendix_biortho}
\mat{\rho}_{{\mathrm{imp,bath}_1}}\mat{U}_{{\mathrm{bath}_1,\mathrm{rest}}}+\mat{\rho}_{{\mathrm{imp},\mathrm{rest}}}\mat{U}_{{\mathrm{rest},\mathrm{rest}}}=\mat{0}.
\end{equation}
This equation has the form of a bi-orthogonality condition and can be solved with a modified Gram-Schmidt procedure. A practical algorithm is provided in the supplemental material at [URL will be inserted by publisher].

This scheme can now be repeated for the submatrix 
\begin{eqnarray}
  \begin{pmatrix} \tilde{\mat{ \rho}}_{{\mathrm{bath}_1,\mathrm{bath}_1}} & \tilde{\mat{ \rho}}_{{\mathrm{bath}_1,\mathrm{rest}}} \\ \tilde{\mat{ \rho}}_{{\mathrm{bath}_1,\mathrm{rest}}}^\dagger & \tilde{\mat{ \rho}}_{{\mathrm{rest},\mathrm{rest}}} \end{pmatrix},
\end{eqnarray}
where we consider $\mathrm{bath}_1$ as the new impurity. Applied iteratively, this yields a sequence of unitary transformations that together form the full transformation $\mat{U}_\mathrm{bath}$.

The unitary transformation $\mat{U}_\mathrm{bath}$ will directly depend on the one-particle density matrix. Thus we named it the \textit{adaptive} cluster approximation (ACA).

In the banded form, Eq.~\eqref{eq:multi_level_trafo} , there are several bath levels: The innermost bath with the density matrix $\tilde {\mat{\rho}}_{\mathrm{bath}_1,\mathrm{bath}_1}$, the second-level bath with $\tilde {\mat{\rho}}_{\mathrm{bath}_2,\mathrm{bath}_2}$ and so on. The interacting one-particle states are only coupled to the innermost bath ($\mathrm{bath}_1$) via $\tilde {\mat{\rho}}_{\mathrm{imp},\mathrm{bath}_1}$ but not to other higher-level baths. In turn, the innermost bath only couples to the second-level bath ($\mathrm{bath}_2$) via $\tilde {\mat{\rho}}_{\mathrm{bath}_1,\mathrm{bath}_2}$ and so on. 
The  lower bound for the  number $N_{\mathrm{bath}_1}$ of orbitals in the innermost bath ($\mathrm{bath}_1$) is less or equal to the number $N_{\mathrm{imp}}$ of interacting orbitals and the  lower bound for the number   $N_{\mathrm{bath}_n}$ of orbitals in the nth-level bath ($\mathrm{bath}_n$) is less or equal to the number $N_{\mathrm{bath}_{n-1}}$ of orbitals in the (n-1)-th level bath ($\mathrm{bath}_{n-1}$). 
The proof of this relation is provided in the supplemental material at  [URL will be inserted by publisher].
 The goal is to have as few orbitals per bath level as possible.  

\subsection{\label{sec:adaptive_cluster_approximations_trunc}Truncation of the bath basis states}
Based on the banded form of the one-particle reduced density matrix in Eq.~\eqref{eq:multi_level_trafo}, we can set up a sequence of approximations: If we neglect the coupling density matrix 
$\tilde{\mat \rho}_{\mathrm{bath}_1,\mathrm{bath}_2}$ we obtain the approximate density matrix $\tilde{\mat{\rho}}_{M=1}\approx \tilde {\mat{\rho}}$ with a block-diagonal form
\begin{equation}
  \tilde {\mat{\rho}}_{M=1}=\left(
    \begin{array}{cc|ccccc}
      \mat{\rho}_{\mathrm{imp},\mathrm{imp}} & \tilde {\mat{\rho}}_{\mathrm{imp},\mathrm{bath}_1} & \mat{0} & \dots \\  
      \tilde {\mat{\rho}}_{\mathrm{imp},\mathrm{bath}_1}^\dagger & \tilde {\mat{\rho}}_{\mathrm{bath}_1,\mathrm{bath}_1} & \mat{0} & \dots  \\ \hline
      \mat{0} & \mat{0}&\tilde {\mat{\rho}}_{\mathrm{bath}_2,\mathrm{bath}_2} & \dots \\
      \vdots & \vdots & \vdots & \ddots \\
    \end{array}
  \right).
\label{eq:ACA_M1}
\end{equation}

The neglect of the coupling between two bath levels $\mathrm{bath}_n$ and $\mathrm{bath}_{n+1}$ is identical to a truncation of all bath levels beyond $\mathrm{bath}_n$ for the purpose of evaluating the density-matrix functional. This is the case because the interaction does not act on the bath.

This yields the adaptive cluster approximation ACA with one effective bath level (ACA(M=1)) by approximating the exact density-matrix functional $F^{\hat W}[\mat{\rho}]$ with
\begin{equation}
  F^{\hat W}_{\mathrm{ACA(M=1)}}[{\mat{\rho}}]:=F^{\hat W}\left [
    \begin{pmatrix}
      \mat{\rho}_{\mathrm{imp},\mathrm{imp}} & \tilde {\mat{\rho}}_{\mathrm{imp},\mathrm{bath}_1}  \\  
      \tilde {\mat{\rho}}_{\mathrm{imp},\mathrm{bath}_1}^\dagger & \tilde {\mat{\rho}}_{\mathrm{bath}_1,\mathrm{bath}_1} 
    \end{pmatrix}\right].
\end{equation}
This approximate density-matrix functional is then used in Eq.~\eqref{eq:omega_min} to obtain the approximate ground-state one-particle reduced density matrix and ground-state energy.
The ACA(M=1) requires only the density-matrix functional for $N_\mathrm{imp}+N_{\mathrm{bath}_1}\leq 2 N_\mathrm{imp}$ one-particle states instead of the original $N_\chi=N_\mathrm{imp}+N_{\mathrm{bath}}$ one-particle states of the full system.

The ACA(M) is obtained by neglecting the coupling between the M-th level and (M+1)-th level bath $\tilde \rho_{\mathrm{bath}_M,\mathrm{bath}_{M+1}}$ instead of $\tilde \rho_{\mathrm{bath}_1,\mathrm{bath}_2}$ (M=1). By increasing $M$ we can systematically converge the approximation.
The size $N_M$ of the one-particle basis treated explicitely in the ACA(M) is
\begin{equation}
  \label{eq:aca_general_number}
    N_M=N_\mathrm{imp}+\sum_{i=1}^M N_{\mathrm{bath}_i}\leq (M+1)N_\mathrm{imp}.
\end{equation} 
In order to judge the quality of the transformation we introduce the discarded weight which measures the deviation from the block-diagonal form. The discarded weight $\sigma_{M}(\tilde{\mat{\rho}})$ is defined as the sum of the absolute values of the neglected coupling density matrix $\tilde{\mat{ \rho}}_{\mathrm{bath}_M,\mathrm{bath}_{M+1}}$:
\begin{eqnarray}
  \label{eq:discarded_weight}
  \sigma_{M}(\tilde{ \mat{\rho}})&=&\sum_{\alpha,\beta} |(\tilde \rho_{\mathrm{bath}_M,\mathrm{bath}_{M+1}})_{\alpha\beta}|.
\end{eqnarray}
A block-diagonal density matrix as Eq.~\eqref{eq:block_rho0} has a vanishing discarded weight $\sigma_{M=0}(\mat{\rho})=0$. The smaller discarded weight the smaller the deviation introduced by the adaptive cluster approximation.  The discarded weight is not a monotonically decreasing function of $M$. However, due to the growing distance from the impurity with increasing bath level $M$ the impact of a finite discarded weight on the ACA is strongly reduced. In applications of the ACA we have not yet seen a case where an increase of discarded weight with increasing bath level $M$, that is $\sigma_{M+1}(\tilde \rho)>\sigma_M(\tilde \rho)$, has caused an increase in the deviation of the density-matrix functional from the exact value, $|F^{\hat W}_{\mathrm{ACA(M)}}[\mat{\rho}]-F^{\hat W}[\mat \rho]|$. The numerical evidence indicates that the error of the ACA is a monotonically decreasing function of $M$.

\subsection{\label{sec:exact_limits}Exact limits}
The ACA can be shown to be exact in a number of limiting cases:
\begin{itemize}
\item It is trivially exact in the non-interacting limit, because the density-matrix functional vanishes in this case.

\item For a single-site impurity ($N_\mathrm{imp}=2$), the ACA with one effective bath site ($M=1$) inherits the exact limits from the related two-level approximation, that have been proven by T\"ows et al.\cite{toews11_prb83_235101}: It becomes exact for a single-impurity Anderson model in both, the limit of a vanishing bath bandwidth and the limit of widely separated bath levels.

\item Furthermore, the ACA(M) is exact if the transformed density matrix $\tilde{\mat{ \rho}}=\mat{U}^\dagger{\mat{\rho}}\mat{U}$, Eq.~\eqref{eq:multi_level_trafo}, is in a block-diagonal form with one block of the size $N_\mathrm{imp}+\sum_{i=1}^M N_{\mathrm{bath}_i}$ and one with the remaining states. 

\item If ACA(M) is exact, then also ACA(M+1) is exact.

\item If the eigenvalue spectrum of the bath density matrix $\mat{\rho}_{\mathrm{bath,\mathrm{bath}}}$ in Eq.~\eqref{eq:rho_as_blocks} consist of $N$ distinct values with a $n_j$-fold degeneracy each, there is a number of bath states $N_B\leq\sum_{j=1}^N \mathrm{min}(n_j,N_\mathrm{imp})$ such that the transformed density matrix obtains a block-diagonal form with one block of the size $N_\mathrm{imp}+N_B$ containing the impurity. Thus, the ACA(N) is exact in the case. 
The proof of this relation is provide in the supplemental material at [URL will be inserted by publisher].
\end{itemize}

\subsection{\label{sec:relation}Related methods}
Methods that employ  transformations or   truncations of one-particle basis sets are omnipresent in the field of quantum chemistry and solid state physics.

The transformation of the one-particle reduced density matrix to a banded matrix within the ACA can be seen as a transformation to a quasi one-dimensional system. Unitary transformations of the one-particle basis to create a quasi one-dimensional system have been used to transform impurity problems with arbitrary bath geometries so that they can be treated with DMRG and related methods\cite{buesser13_prb88_245113,shirakawa14_prb90_195109}. 
Unitary transformations of the bath in impurity problems have also been used to express the ground state with a small number of Slater determinants\cite{PhysRevB.88.035123,PhysRevB.90.085102} or to set up a smaller variationally optimized effective model\cite{maltevared}. These methods differ from our approach in that they bring the one-particle Hamiltonian rather than the reduced density matrix to a specific shape.

In context of rDMFT, an approach similar to the ACA has been suggested and used by T\"ows et al. within the two-level approximation\cite{toews11_prb83_235101,toews12_prb86_245123} for the single-impurity Anderson\cite{anderson61_pr124_41} model.
T\"ows et al. derived the analytical dependence of the density-matrix functional with respect to the density matrix for a simple Anderson problem, i.e. an impurity site and one bath site. To apply this analytical form to more general single-impurity Anderson models, they introduced a unitary transformation of the bath states, so that only two of the transformed bath states have finite density matrix elements with the impurity site. All other bath states have been neglected in the evaluation of the density-matrix functional. 
Although constructed in a different way, the transformation of T\"ows et al. in the case of two interacting spin-orbitals ($N_\mathrm{imp}=2$) is equivalent to the first transformation step in our construction scheme (see supplemental material at  [URL will be inserted by publisher]).
T\"ows et al. used the exact density-matrix functional for the effective two-site problem (impurity site and first-level effective bath site) in case of a non-spin-polarized density matrix\cite{toews11_prb83_235101}. For the spin-polarized case\cite{toews12_prb86_245123} they employed additional approximations. 
In contrast, in this paper we calculate the density-matrix functional for the truncated density matrix on the fly via a constrained optimization scheme without additional approximations.
The ACA(M) is the extension of two-level approximation to an arbitrary number of effective bath states/levels, arbitrary impurity sizes and multi-band interactions.

\subsection{\label{sec:correction}Correction using parametrized functionals}
The minimization of the energy given in Eq.~\eqref{eq:omega_min} with the adaptive cluster approximation can produce density matrices with a discarded weight $\sigma_{M}(\tilde{\mat{ \rho}})$ that is much larger than that for the transformed exact ground-state density matrix.
The reason is, that the truncated off-diagonal density matrix elements $\tilde{\mat{ \rho}}_{\mathrm{bath}_{M}\mathrm{bath}_{M+1}}$ do not influence the density-matrix functional within the ACA(M). As a consequence during the minimization over the density matrix they can differ strongly from the exact solution if this reduces the one-particle energy in Eq.~\eqref{eq:omega_min}. Our applications to model systems indicate that this becomes relevant only at very low truncation levels.

Nevertheless, in order to cope with this problem we add a correction term $\Delta F^{\hat{W}}_{\approx}\left[\mat{\rho}\right]$
\begin{equation}
  \Delta F^{\hat{W}}_{\approx}\left[\mat{\rho}\right]=F^{\hat{W}}_{\approx}\left[\tilde{ \mat{\rho}}\right]-F^{\hat{W}}_{\approx}\left[ \tilde{\mat{ \rho}}_M\right],
\label{eq:adaptive_cluster_M_corr2}
\end{equation}
that is the difference of an approximate density-matrix functional for the full transformed density matrix $\tilde {\mat \rho}$, Eq.~\eqref{eq:multi_level_trafo}, and the truncated density matrix $\tilde{\mat{ \rho}}_M$, where the coupling density-matrix elements $\tilde{ \mat \rho}_{\mathrm{bath}_M,\mathrm{bath}_{M+1}}$  has   been neglected. Thus, the corrected ACA (cACA) is defined by 
\begin{equation}
  F^{\hat{W}}_{\mathrm{cACA(M)}}\left[\mat{\rho}\right]=F^{\hat{W}}_{\mathrm{ACA(M)}}\left[\mat{\rho}\right]+\Delta F^{\hat{W}}_{\approx}\left[\mat{\rho}\right].
\label{eq:adaptive_cluster_M_corr}
\end{equation}
The correction $\Delta F^{\hat{W}}_{\approx}\left[\mat{\rho}\right]$ is expressed with the help of a density-matrix functional $F^{\hat{W}}_{\approx}[\mat{\rho}]$, that should be easy to evaluate. Suitable are for example parametrized approximations in an analytical form.

In this work, we choose the M\"uller density-matrix functional\cite{mueller84_pl105A_446} as the correction functional $F^{\hat{W}}_{\approx}[\mat{\rho}]$ in Eq.~\eqref{eq:adaptive_cluster_M_corr2}. The M\"uller functional is defined by the analytical expression
\begin{equation}
  \label{eq:mueller}
  F^{\hat{W}}_{\approx}[\mat{\rho}]=\frac{1}{2}\sum_{a,b,c,d} U_{a,b,c,d} \left[ \rho_{d,a} \rho_{c,b}-(\mat \rho^\frac{1}{2})_{ca}(\mat \rho^\frac{1}{2})_{db}\right],
\end{equation}
where $f_n$ denote the occupations, $|\phi_{n}\rangle$ the natural orbitals, $|\chi_a\rangle$ a orthonormal one-particle basis set and $U_{a,b,d,c}$ interaction matrix elements. 

A non-vanishing effect of the correction requires that the derivatives of $F^{\hat{W}}_{\approx}\left[\tilde{\mat{ \rho}}\right]$ with respect to the off-diagonal coupling matrices $\tilde {\mat{\rho}}_{\mathrm{bath}_{n}\mathrm{bath}_{n+1}}$  in Eq.~\eqref{eq:multi_level_trafo} don't vanish by construction. 
Otherwise for example in the case of ACA(M=1), that is when neglecting $\tilde {\mat{\rho}}_{\mathrm{bath}_1,\mathrm{bath}_2}$, we would get
\begin{equation}
  \frac{\partial \Delta F^{\hat{W}}_{\approx}\left[\mat{\rho}\right]}{\partial \tilde {\mat{\rho}}_{\mathrm{bath}_1,\mathrm{bath}_2}}=\frac{\partial F^{\hat{W}}_{\approx}\left[\tilde{ \mat{\rho}}\right]}{\partial \tilde {\mat{\rho}}_{\mathrm{bath}_1,\mathrm{bath}_2}}-\frac{\partial F^{\hat{W}}_{\approx}\left[ \tilde{\mat{ \rho}}_M\right]}{\partial \tilde {\mat{\rho}}_{\mathrm{bath}_1,\mathrm{bath}_2}}=0.
\end{equation}
Thus the correction term $\Delta F^{\hat{W}}_{\approx}\left[\mat{\rho}\right]$ would yield a vanishing force on the neglected density matrix elements and not fulfill its purpose of preventing an unphysical increase of the neglected density matrix elements.
For example this condition is not fulfilled for the Hartree-Fock approximation, because in this approximation the value of the functional,
\begin{equation}
  \label{eq:HF}
  F^{\hat{W}}_{\mathrm{HF}}[\mat{\rho}]=\frac{1}{2}\sum_{abcd} U_{abdc} (\rho_{da} \rho_{cb}-\rho_{ca}\rho_{db}),
\end{equation}
is independent of density matrix elements of non-interacting states.

\subsection{\label{sec:beyondsiam}Beyond single-impurity Anderson models}
The extension of the ACA from single impurity Anderson models to lattice models is straightforward when we introduce the local approximation of the density-matrix functional\cite{bloechl11_prb84_205101}:
For a Hubbard-like interaction, that can be written as as sum over local terms,
\begin{equation}
  \hat W=\sum_i \hat W_i,
\end{equation}
we can approximate the density-matrix functional as sum over local density-matrix functionals:
\begin{equation}
  F^{\hat{W}}[\mat{\rho}]\approx \sum_i F^{\hat{W_i}}[\mat{\rho}].
\end{equation}
The individual density-matrix functionals $F^{\hat{W_i}}[\mat{\rho}]$ now have the form of a single-impurity Anderson model and can be treated with the adaptive cluster approximation. 
Investigations for model lattice systems are is progress.

The purpose of examinations of the ACA on model systems such as the SIAM is to benchmark it for systems that are well understood and to learn about it's strengths and weaknesses. We have constructed the ACA in context of hybrid theories that combine DFT and rDMFT~\cite{PhysRevA.81.052511,PhysRevA.82.052502,bloechl11_prb84_205101,epjstrdmft}. The main idea of such a hybrid approach is to treat the degrees of freedom responsible for strong non-dynamical correlation with an rDMFT-functional and the dynamical correlation with an existing density functional. For materials such as transition-metal oxides it is known that the local atomic physics of the transition metal multiplets is responsible for the non-dynamical correlation, which causes DFT with approximate semi-local or local functionals to fail. 
Despite successes\cite{sharma08_prb78_201103,sharma13_prl110_116403,shinohara1} of the simple power functional\cite{sharma08_prb78_201103} some severe failures of widely used functionals have also been demonstrated for model systems~\cite{kamil16_prb93_085141,PhysRevA.93.042511}. Thus there is a need for better functionals and especially functionals, that can be systematically improved. 
We see the future applicability of the adaptive cluster approximation in the context of such an advanced functional that contains an internal minimization problem and an unfavorable scaling.
The ACA reduces the number of the non-interacting orbitals when using implicit functionals with the local approximation in DFT+rDMFT\cite{bloechl11_prb84_205101,epjstrdmft}.

\section{\label{sec:numerical_methods}Numerical methods}
\subsection{\label{sec:numerical_methods_0}Exact ground-state energies and many-particle wave functions}
To benchmark the performance of the adaptive cluster approximation for a single-impurity Anderson model, we have calculated the exact ground-state energy and many-particle wave function for this system at zero temperature with an exact diagonalization algorithm based on the Jacobi-Davidson method implemented in Jadamilu\cite{bollhoefer07_cpc177_951}.

 For the single-impurity Anderson model with a large bath in section~\ref{sec:siam_beyond} we have used the matrix-product-state-DMRG\cite{white92,white93,SCHOLLWOCK201196} code ITensor\cite{itensor} and employed particle number conservation and spin-rotation symmetry\cite{PhysRevB.83.115125}. A measure for the accuracy of the results from a DMRG calculation is the maximal truncation error $\epsilon=\max_i \epsilon_i$. 
The truncation error $\epsilon_i$ of an individual bond during the two-site DMRG procedure is defined in ITensor as
\begin{equation}
\epsilon_i=\frac{\sum_{n\in\mathrm{neglected}} \lambda_{i,n}^2}{\sum_{n} \lambda_{i,n}^2},
\end{equation}
where $\lambda_{i,n}$ are the singular values. The number of singular values that are not neglected is called the bond dimension. The error introduced into the many-particle wave function $|\psi\rangle$ from the truncation is in the worst case\cite{PhysRevB.73.094423,SCHOLLWOCK201196}
\begin{equation}
  \parallel |\psi\rangle-|\psi_\mathrm{trunc}\rangle\parallel_2^2\leq 2 \sum_i \epsilon_i.
\end{equation}
The systems studied in section~\ref{sec:siam_beyond_all} are Anderson models with single- or multi-site impurities with hopping from every impurity site to every bath site. Thus the one-particle Hamiltonian is very non-local and the entanglement entropy of the ground state is expected to be high in this one-particle basis. In order to make the Hamiltonian more local we have transformed the matrix elements one-particle Hamiltonian to a banded form like B\"usser et al.\cite{buesser13_prb88_245113} with the same algorithm that we use for the one-particle reduced density matrix in Eq.~\eqref{eq:multi_level_trafo}. The resulting one-particle Hamiltonian has a structure that is similar to the one-particle Hamiltonian of a two-leg Hubbard ladder. For a single-site-impurity model with 40 sites, the interaction strength $U=1\ \mathrm{eV}$ and the parameters described in detail in section~\ref{sec:siam_beyond_all}, we need a bond dimension of about 430 to reach a truncation error of $10^{-7}$. This small bond dimension is numerically straight forward but we expect much higher entanglement entropies and thus bond dimensions for models with multi-site impurities as the structure of the one-particle Hamiltonian becomes similar to a $n$-leg Hubbard ladder, where the number $n$ of legs is twice the number of sites in the impurity.  

Since we have to calculate the one-particle reduced density matrix for the whole system anyway, we used this information to construct a more suitable transformation of the one-particle basis. 
Here it is interesting to note that the matrix elements of the one-particle reduced density matrix are in principle only correlation functions and that matrix product states can't represent algebraically decaying correlation functions\cite{SCHOLLWOCK201196} efficiently. Thus we apply the following computational scheme: We first perform a cheap DMRG calculation with a small bond dimension and the one-particle Hamiltonian in a banded form. This gives us a rough estimate of the one-particle reduced density matrix of the system. Then we transform the one-particle basis such that the estimated one-particle reduced density matrix has a banded form as in Eq.~\eqref{eq:multi_level_trafo} making the correlation functions $\langle \hat c_\alpha^\dagger \hat c_\beta\rangle$ very short range. We then transform the matrix elements of the one-particle Hamiltonian into the same basis and perform the main DMRG calculation. For the 40-site SIAM with a single-site impurity mentioned earlier the final DMRG calculation only needs a fifth of the bond dimension, 80 instead of 430, to reach a maximal truncation error of $10^{-7}$.  
 
\subsection{\label{sec:numerical_methods_1}Minimization over the one-particle reduced density matrix}
The minimization over the ensemble-representable one-particle reduced density matrix in Eq.~\eqref{eq:omega_min} is performed using a Car-Parrinello-like\cite{car85_prl55_2471} constrained minimization. For that purpose the density matrix is written in its spectral representation as
\begin{eqnarray}
  \label{eq:rho_occ_natorb}
  \rho_{\alpha\beta} &=&\sum_{i=1}^{N_\chi}f_i \phi_{i,\alpha} \phi_{i,\beta}^*
\end{eqnarray}
with the occupations $f_i$ and the normalized eigenvectors $\mat{\phi}_{i}$, which are called natural orbitals\cite{loewdin55_pr97_1474}. The ensemble representability of the  density matrix requires the occupations to be between 0 and 1. This condition is satisfied by expressing the occupations as $f_i=[1-\cos(x_i)]/2$ with unconstrained variables $x_i$. Using the set of $x_i$ and the natural orbitals as dynamical quantities, a fictitious Lagrangian for the calculation of the energy can be set up in the form
\begin{eqnarray}
  \mathcal{L}(\{x_i\},&\{\dot x_i\}&,\{\mat{\phi}_i\},\{\dot {\mat{\phi}}_i\})= \frac{1}{2}\sum_{i=1}^{N_\chi} m_x \dot x_i^2 \nonumber\\
&+&\sum_{i=1}^{N_\chi} f(x_i)m_\phi |\dot {\mat{\phi}}_i|^2 \nonumber\\
             &-&\mathrm{Tr}[\mat{\rho}\mat{ h}]-F^{\hat W}[\mat{\rho}]+\mu\left(\mathrm{Tr}[\mat{\rho}]-N\right)
\nonumber\\
             &+&\sum_{i,j}\Lambda_{i,j} \left(\mat{\phi}_i^* \cdot \mat{\phi}_j-\delta_{i,j}\right)
\end{eqnarray}
where $\mat{\rho}$ is given in terms of the occupations and natural orbitals by Eq.~\eqref{eq:rho_occ_natorb}. $\mu$ and $\Lambda_{i,j}$ are Lagrange multipliers for the particle number constraint and orthonormality.

Starting from a random initial guess, the Euler-Lagrange equations are integrated using the Verlet algorithm\cite{PhysRev.159.98}. 
The particle number constraint and  the orthonormality constraint of the natural orbitals are enforced in every time step of the integration with the help of the corresponding Lagrange multipliers\cite{ryckaert77_jcompphys23_327}. 

A minimum of the potential energy
\begin{eqnarray}
  \mathrm{Tr}[\mat{\rho}\mat{ h}]+F^{\hat W}[\mat{\rho}]
\end{eqnarray}
 of the fictitious Lagrangian with respect to the constraints is obtained by including an additional friction term.
The convergence criterion for the numerical minimization of the total energy given by Eq.~\eqref{eq:omega_min} is chosen as $10^{-4}t$, where $t$ is the hopping parameter. 
The convergence is verified by propagating the Car-Parriniello dynamics without friction for a large number of steps and checking that the energy stays within a window defined by the given convergence criterion.
The search space includes density matrices with broken spin-symmetry and non-collinear density matrices. 

\subsection{\label{sec:numerical_methods_2}Evaluation of the exact density-matrix functional}
We explore the performance of the (corrected) adaptive cluster approximation in the zero-temperature limit. We evaluate the density-matrix functional numerically by solving Eq.~\eqref{eq:rDMF2}. 
The many-particle wave function is represented as a superposition of Slater determinants with variable coefficents. A complete set of Slater determinants is used in the present study. 
Details for the practical solution of the constrained minimization problem are given in the supplemental material at  [URL will be inserted by publisher]. 

\section{\label{sec:siam}Benchmark results for the single-impurity Anderson model}
\subsection{\label{sec:siam_model}Definition of the model}
To investigate the properties of the adaptive cluster approximation, we have chosen the same finite single-impurity Anderson\cite{anderson61_pr124_41} model as in the first publication of the two-level approximation of T\"ows and Pastor\cite{toews11_prb83_235101}.
It consists of one impurity site , $L_\mathrm{imp}=1$,   with a local density-density interaction and a ring of $L_{\mathrm{bath}}$  non-interacting bath-sites with nearest-neighbor hopping. Electrons on the impurity site can hop directly to only one of the bath sites. 
The Hamiltonian can be divided into three parts:
\begin{equation}
  \label{eq:SIAM_H}
  \hat H=\hat H_{\mathrm{imp}}+\hat H_{\mathrm{bath}}+\hat H_{\mathrm{hyb}}\;.
\end{equation}
The impurity Hamiltonian $\hat H_{\mathrm{imp}}$ contains the impurity on-site energy $\epsilon_f$ and a local two-particle interaction $\hat W=U \hat n_{f,\uparrow}\hat n_{f,\downarrow}$ with the interaction parameter $U$ 
\begin{equation}
  \label{eq:SIAM_H_imp}
  \hat H_{\mathrm{imp}}=\sum_{\sigma\in \{\uparrow,\downarrow\}} \epsilon_f \hat f^\dagger_{\sigma}\hat f_{\sigma}+\hat W\;.
\end{equation}
  $\hat f^\dagger_{\sigma}$ ($\hat f_{\sigma}$) denotes the creation (annihilation) operator for an electron in the spin-state $\sigma \in \{\uparrow,\downarrow\}$ in the impurity orbital. The number operator for the impurity orbital is $\hat n_{f,\sigma}=\hat f^\dagger_{\sigma}\hat f_{\sigma}$.
The one-particle Hamiltonian of the bath, i.e. the non-interacting ring  with hopping parameter $t>0$ , has the form
\begin{equation}
  \label{eq:SIAM_H_bath_real}
  \hat H_{\mathrm{bath}}=-t\sum_{\langle i,j\rangle}\sum_{\sigma\in \{\uparrow,\downarrow\}} \hat c^\dagger_{i,\sigma}\hat c_{j,\sigma}+h.c.,
\end{equation}
where $\hat c_{i,\sigma}$ ($\hat c^\dagger_{i,\sigma}$) denotes the creation (annihilation) operator for a state at the bath site $i$ with spin $\sigma$. The notation $\langle i,j\rangle$ restricts the summation to pairs of nearest-neighboring bath sites.
The hybridization Hamiltonian $\hat H_{\mathrm{hyb}}$, that describes the hopping between impurity and bath, can be written as
\begin{equation}
  \label{eq:SIAM_H_hyb_real}
  \hat H_{\mathrm{hyb}}=V \sum_{\sigma\in \{\uparrow,\downarrow\}}\left(\hat f^\dagger_{\sigma}\hat c_{1,\sigma}+\hat c^\dagger_{1,\sigma}\hat f_{\sigma}\right)
\end{equation}
with the hybridization parameter $V$.
The energy eigenvalues of the one-particle Hamiltonian of the non-interacting bath in Eq.~\eqref{eq:SIAM_H_bath_real} are
\begin{equation}
  \label{eq:epsilon_k}
\epsilon_k=-2t\cos\left (\frac{2\pi k}{L_{\mathrm{bath}}} \right),
\end{equation}
with $k\in \{0,...,L_{\mathrm{bath}}-1\}$.  The lowest one-particle energy level $-2t$ has a multiplicity of two and all other energy levels have a multiplicity of four. 

\subsection{\label{sec:siam_Udep}Interaction-strength dependence}
In this section, we compare exact results to the (corrected) adaptive cluster approximation to evaluate its performance for this system. We show exact zero-temperature results for the total ground-state energy $E_\text{exact}=\langle\psi_\text{exact}|\hat H|\psi_\text{exact} \rangle$, interaction energy $W_\text{exact}=\langle\psi_\text{exact}|\hat W |\psi_\text{exact} \rangle$ and impurity occupation $n_{f,\text{exact}}=\langle\psi_\text{exact}|\hat n_{f,\uparrow}+\hat n_{f,\downarrow}|\psi_\text{exact} \rangle$ at half filling, i.e. with $L_{\mathrm{bath}}+1$ electrons, and briefly discuss the physical background. 
The interaction-strength dependence is investigated in this section. The impurity on-site-energy dependence and bandwidth dependence can found in the supplemental material at  [URL will be inserted by publisher].

\subsubsection{\label{sec:siam_Udep_exact}Exact results}

\begin{figure}[htb]
 \includegraphics[width=\linewidth,height=!]{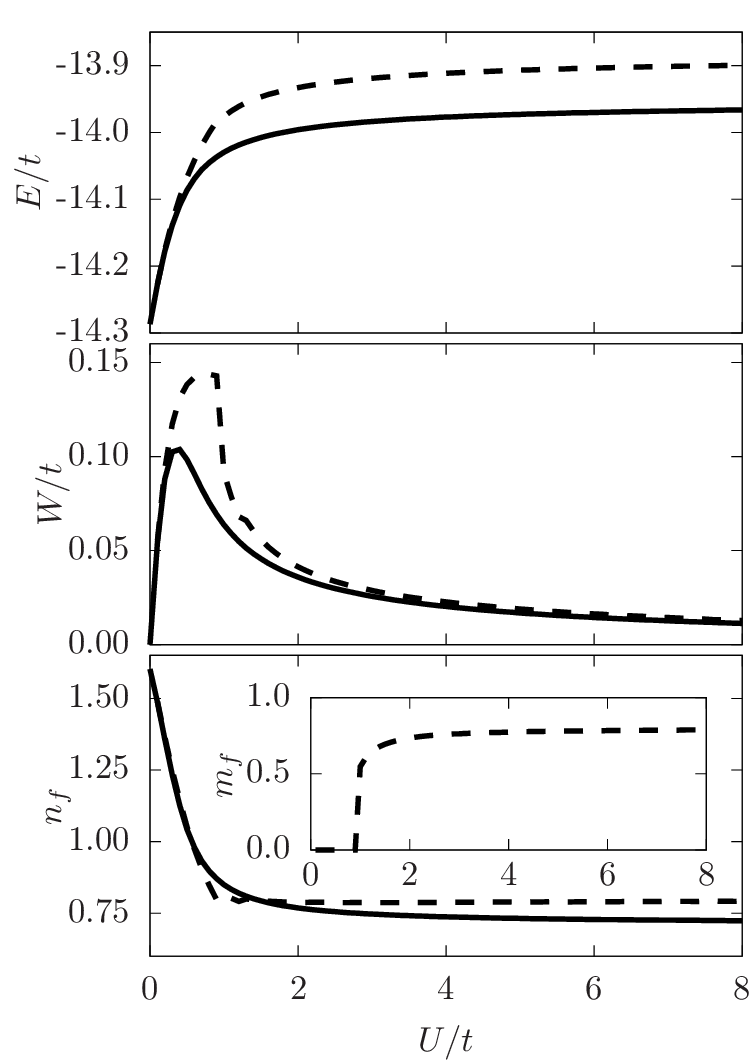}
  \caption{\label{fig:SIAMTLAexactUdep} Ground-state energy $E$, interaction energy $W$ and impurity occupation $n_{f}$ of the exact ground state (solid lines) and the unrestricted Hartree-Fock ground state (dashed lines) of the Hamiltonian defined Eq.~\eqref{eq:SIAM_H}-\eqref{eq:SIAM_H_hyb_real} with $L_{\mathrm{bath}}=11$, $\epsilon_f=0$, $t>0$, $V/t=0.4$ at half filling. 
   The inset in the third graph shows the magnetic moment $m_f=n_{f,\text{exact},\uparrow}-n_{f,\text{exact},\downarrow}$ of the impurity within the unrestricted Hartree-Fock approximation.
 }
\end{figure}

The influence of the interaction strength $U/t$ on the exact and the Hartree-Fock ground state is shown in Fig.~\ref{fig:SIAMTLAexactUdep}.  The parameters were $L_{\mathrm{bath}}=11$, $\epsilon_f=0$, $t>0$, $V/t=0.4$ and $U/t\in [0,8]$. The SIAM is half-filled, i.e. the particle number is fixed to $N_e=L_{\mathrm{bath}}+1=12$. Consequently, the impurity can be occupied with up to two electrons, while the bath can contain between 10 and 12 electrons. 
The energy levels of the bath are given in Eq.~\eqref{eq:epsilon_k}.

The Fermi level $\epsilon_{F,\mathrm{bath}}$ of the bath for a particle number $N$ is defined as
\begin{equation}
  \epsilon_{F,\mathrm{bath}}=\lim_{n\rightarrow N+} \frac{\partial E_{\mathrm{bath}}(n)}{\partial n},
\end{equation}
where $E_{\mathrm{bath}}(n)$ is the energy of the non-interacting bath with $n$ electrons.  
The bath Fermi level $\epsilon_{F,\mathrm{bath}}$ for the model under investigation is given by the one-particle energy of the eleventh and twelfth energy level of the non-interacting bath,
\begin{equation}
  \epsilon_{F,\mathrm{bath}}=-2t\cos(2\pi\cdot 3/11)\approx 0.28t>0.
\end{equation}
Because the impurity level $\epsilon_f=0$ is chosen to lie below the bath-Fermi level $\epsilon_{F,\mathrm{bath}}>0$ the impurity is more than half-filled, i.e. $n_f>1$, in the non-interacting limit $U=0$. Due to the finite impurity-bath hybridization the impurity is not completely filled except for a completely filled bath.
When increasing the interaction strength from $U=0$, electrons are transferred from the impurity to the bath, because the interaction penalizes the double occupancy $\langle\hat n_{f,\uparrow}\hat n_{f,\downarrow}\rangle$ on the impurity. For small interaction strengths we have the simple quadratic relation $W\approx Un_f^2/4$ between the interaction energy $W$ and the impurity occupation. This relation follows directly from the Hartree-Fock approximation of the density-matrix functional for a non-magnetic one-particle reduced density matrix. 

As anticipated, the spin-unrestricted Hartree-Fock approximation agrees well with the exact result for small interaction strengths and yields an upper bound to the exact ground-state energy. However, as shown in the inset of Fig.~\ref{fig:SIAMTLAexactUdep}, the HF solution undergoes a transition to a qualitatively incorrect ground state at $U\approx t$ with a finite magnetization $m_f=n_{f,\uparrow}-n_{f,\downarrow}$ on the impurity. 

\subsubsection{\label{sec:siam_Udep_fullmin}ACA-results}
\begin{figure}[htb]
 \includegraphics[width=\linewidth,height=!]{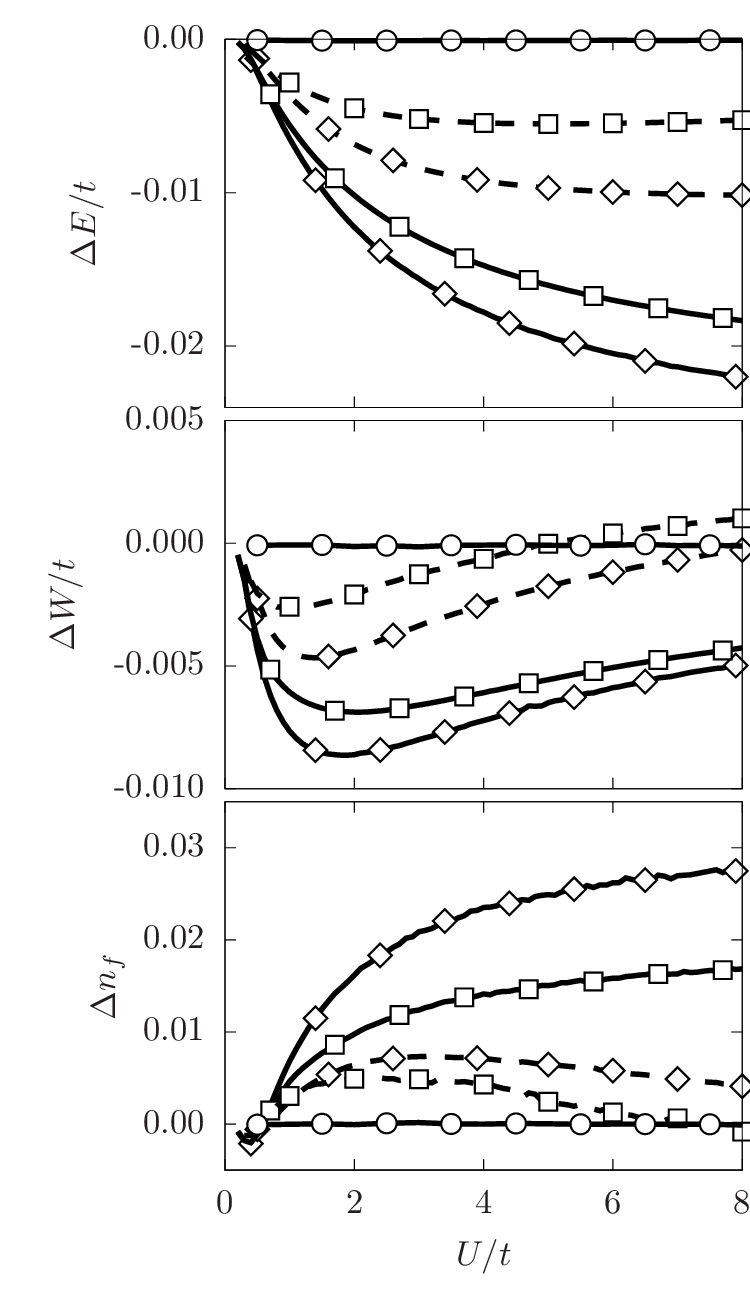}
  \caption{\label{fig:SIAMTLAUdepfullminaca1} Deviation of the ground-state energy $\Delta E=E_{(c)ACA(M)}-E_\text{exact}$, interaction energy $\Delta W=W_{(c)ACA(M)}-W_\text{exact}$ and impurity occupation $\Delta n_{f}=n_{f,(c)ACA(M)}-n_{f,\text{exact}}$ between the ground state within the (corrected) adaptive cluster approximation and the exact ground state. The corresponding exact results are shown in Fig.~\ref{fig:SIAMTLAexactUdep}. Same model as in Fig.~\ref{fig:SIAMTLAexactUdep}. Solid lines correspond to the uncorrected ACA and dashed lines to the M\"uller-corrected ACA results. 
   Truncation with $M=1$ (solid line with diamonds), $M=2$ (solid line with squares) and $M=3$ (solid line with circles). The noise, especially in $\Delta n_f$, is due to the finite convergence criterion for the minimization in Eq.~\eqref{eq:omega_min}.
 }
\end{figure}

In figure~\ref{fig:SIAMTLAUdepfullminaca1}, the deviations of energies and impurity occupations obtained by the ACA from the exact results are shown as function of interaction strength.  We have chosen the number of orbitals per bath level $N_{\mathrm{bath}_i}$ equal to the number of impurity orbitals $N_{\mathrm{imp}}=2$.  

\begin{figure}[htb]
 \includegraphics[width=\linewidth,height=!]{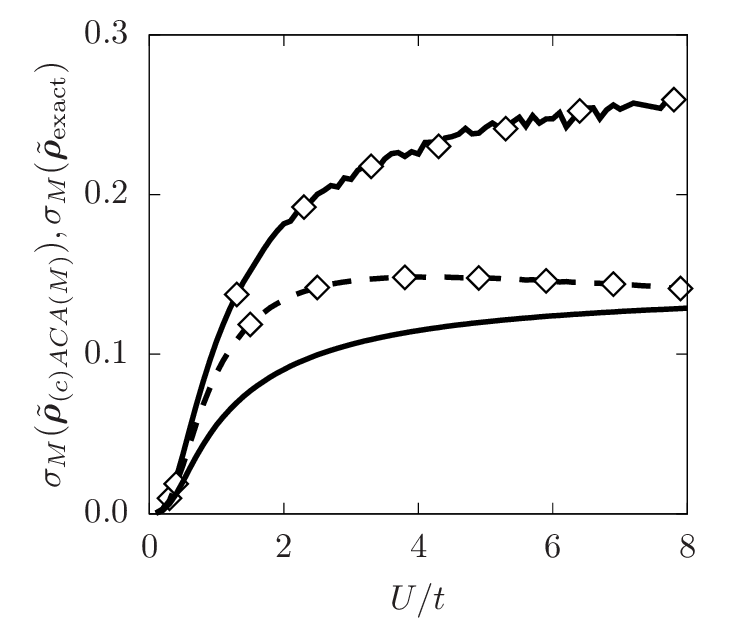}
  \caption{\label{fig:SIAMTLAUdepfullminaca1sigma} Discarded weight $\sigma_{M}$ given by Eq.~\eqref{eq:discarded_weight} of the transformed ground state density matrix $\tilde {\mat{\rho}}_{(c)ACA(M)}$ within the (corrected) ACA and the transformed exact density matrix $\tilde {\mat{\rho}}_{\text{exact}}$. Same model as in Fig.~\ref{fig:SIAMTLAexactUdep}. The ACA-truncation with $M=1$ of the transformed exact density matrix $\tilde{\mat{\rho}}_{\text{exact}}$ (solid line without symbols) is compared the ground state of the uncorrected (solid line with symbols) and corrected (dashed line with symbols) ACA with $M=1$.
 }
\end{figure}

An important success is that, unlike the Hartree-Fock approximation, the ACA does not break the spin-symmetry. This is true even for the lowest truncation level $M=1$. The ground-state density matrix obtained with the ACA agrees well with the exact one. 
The ACA overestimates the impurity occupation $n_f$. This is a consequence of the underestimation of the interaction energy, which effectively reduces the electron repulsion on the impurity.
On the one hand, the uncorrected ACA shares many features with the two-level-approximation of T\"ows et al., notably the large relative error of the interaction energy for large interaction strengths\cite{toews11_prb83_235101}. 
The M\"uller-corrected ACA, on the other hand, greatly improves the results for the entire range of interaction strengths. As intended, the M\"uller correction improves the result by preventing the growth of discarded weight during the density matrix optimization. 
The reduction of the discarded weight by the  M\"uller-corrected ACA   has been discussed in Sec.~\ref{sec:correction} and it is demonstrated in Fig.~\ref{fig:SIAMTLAUdepfullminaca1sigma}.

For large interactions, the M\"uller correction overcorrects the results for interaction energy and impurity occupation, which is a consequence of the well-known\cite{staroverov02_jcp117_2489,lathiotakis08_jcp128_184103,lathiotakis07_prb75_195120} overcorrelation  of M\"uller's functional. The overcorrelation of M\"uller's functional is attenuated in the context of the ACA correction, because it does not affect the dominant contributions as discussed in~\ref{sec:correction}.

\begin{figure}[htb]
 \includegraphics[width=\linewidth,height=!]{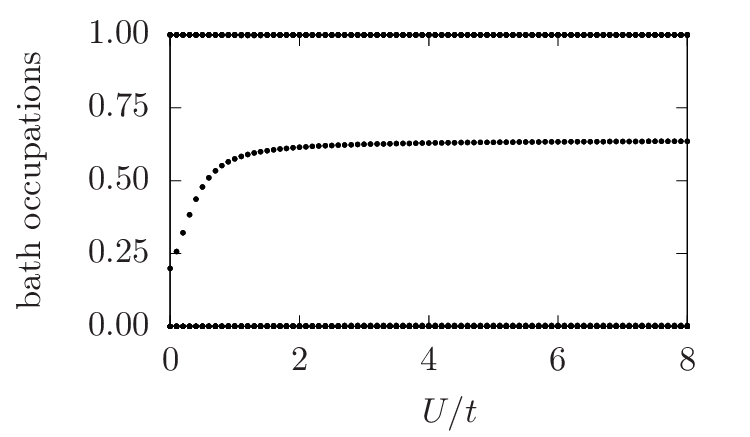}
  \caption{\label{fig:SIAMTLAUdepoccb} Eigenvalue spectrum (occupations) of the bath density matrix $\mat{\rho}_{\text{bath,bath,exact}}$ of the exact ground state. Same model as in Fig.~\ref{fig:SIAMTLAexactUdep}.
 }
\end{figure}

 The ACA with $M=3$, that is three effective bath sites, converges to the exact ground-state energy and ground-state density matrix within the convergence criteria of the numerical minimization procedures used here. This can be understood when consulting the eigenvalue spectrum of the exact bath density matrix $\mat{\rho}_{\text{bath,bath,exact}}$ as shown in Fig.~\ref{fig:SIAMTLAUdepoccb}. There are three ($N=3$) clusters of eigenvalues. Thus as explained in sec.~\ref{sec:exact_limits} the ACA would need at most $N \cdot N_\mathrm{imp}=6=2\cdot M$ effective bath states to be exact if the clusters had exactly degenerate eigenvalues. However the eigenvalues in two of the clusters have a finite spread: There is one set of ten small eigenvalues with values below $4\cdot 10^{-3}$. A second set of ten eigenvalues lies close to 1, i.e. between $1-10^{-4}$ and 1. The third set that contains two degenerate eigenvalues lies between the first two sets. Due to the very small but finite spread of the eigenvalues the discarded weight does not vanish for $M=3$ but has very small finite values between zero and $10^{-4}$ in this parameter range. This implies that the ACA produces an extremely small error of the density-matrix functional that is below the practically feasible convergence criterion of the constrained minimization for the density-matrix functional and not physically relevant. The dependence of the discarded weight for the exact ground state of single-impurity Anderson models with larger baths and multi-orbital impurities is investigated in the subsequent section.

 In conclusion,   the ACA with and without correction describes the interaction-strength dependence of the single-impurity Anderson model very well in both the weakly and the strongly interacting regime. Correlation effects are correctly described without unphysical spin-symmetry breaking.


\section{\label{sec:siam_beyond}Behaviour for larger bath sizes and multi-orbital SIAMs}
\subsection{\label{sec:siam_beyond_model}Definition of the model}
 
The single-impurity Anderson model used in this work to investigate the performance of the ACA seems to have the feature, that at zero temperature the eigenvalue spectrum of the bath density matrix $\mat \rho_{BB}$ has only three clusters of eigenvalues with a very small spread. We numerically found this to be true for the model in Eq.~\eqref{eq:SIAM_H}-\eqref{eq:SIAM_H_hyb_real} in all parameter ranges that we studied except for vanishing bandwidth $t=0$. As the number of clusters is an indicator for the performance of the ACA we study this quantity for SIAMs beyond the one studied in section~\ref{sec:siam_model}. We focus on the dependence on the bath size and number of impurity orbitals. Even though Anderson models have been studied intensely, we found no systematic investigations of the eigenvalue spectrum of the one-particle reduced density matrix.
 
We choose a single-impurity multi-orbital Anderson model with $N_\mathrm{imp}=2L_\mathrm{imp}$ impurity orbitals. $\hat f^\dagger_{i,\sigma}$ ($\hat f_{i,\sigma}$) denote the creation (annihilation) operators for an electron in the spin-state $\sigma \in \{\uparrow,\downarrow\}$ in the impurity site with index $i$. These impurity sites have an on-site energy of $\epsilon_{f,i}$ and a local interaction $\hat W_i=U_i \hat n_{f,i,\uparrow} \hat n_{f,i,\downarrow}$ with $\hat n_{f,i,\sigma}=\hat f^\dagger_{i,\sigma}\hat f_{i,\sigma}$,

\begin{equation}
  \label{eq:SIAM2_H_imp}
  \hat H_{\mathrm{imp}}=\sum_{i=1}^{L_{imp}}\sum_{\sigma\in \{\uparrow,\downarrow\}} \epsilon_{f,i} \hat f^\dagger_{\sigma,i}\hat f_{\sigma,i}+\sum_{i=1}^{L_{imp}} \hat W_i.
\end{equation}
The one-particle Hamiltonian of the $L_\mathrm{bath}$ bath sites,
\begin{equation}
  \label{eq:SIAM2_H_bath}
  \hat H_{\mathrm{bath}}=\sum_{j=1}^{L_\mathrm{bath}} \sum_{\sigma\in \{\uparrow,\downarrow\}} \epsilon_{b,j} \left(\hat c^\dagger_{j,\sigma}\hat c_{j,\sigma}+h.c.\right),
\end{equation}
includes the bath energy levels $\epsilon_{b,j}$ that are uniformly distributed in the interval $[\epsilon_b-w_b/2,\epsilon_b+w_b]$ where $\epsilon_b$ denotes the mean bath energy and $w_b$ the bandwidth of the bath. In the limit of infinitely many bath sites this yields a continuous flat density of states.
The hopping between the impurity sites and the bath sites is described by the hybridization Hamiltonian
\begin{equation}
  \label{eq:SIAM2_H_hyb}
  \hat H_{\mathrm{hyb}}=\sum_{i=1}^{L_\mathrm{imp}} \sum_{j=1}^{L_\mathrm{bath}} V_{i,j} \sum_{\sigma\in \{\uparrow,\downarrow\}}\left(\hat f^\dagger_{i,\sigma}\hat c_{j,\sigma}+h.c.\right).
\end{equation}
The ground state of the total Hamiltonian,
\begin{equation}
  \label{eq:SIAM2_H_tot}
  \hat H=\hat H_{\mathrm{imp}}+\hat H_{\mathrm{bath}}+\hat H_{\mathrm{hyb}},
\end{equation}
at zero temperature is then solved with matrix-product-state-DMRG\cite{white92,white93,SCHOLLWOCK201196}. 

\subsection{\label{sec:siam_beyond_malte}Relation to the SIAMs studied by Sch\"uler et al.}
Single-impurity Anderson models with finite baths of the form in Eq.~\eqref{eq:SIAM2_H_imp}-Eq.~\eqref{eq:SIAM2_H_tot} have been investigated for example by Sch\"uler et al.\cite{maltevared,Schueler2017}. The single-impurity Anderson model used as benchmark cases by Sch\"uler et al. for their variational exact diagonalization method can be represented by the above model with the choices $L_\mathrm{imp}=1$, $L_\mathrm{bath}=6$, $\epsilon_{f,1}=-2\ \mathrm{eV}$, $U_1=4\ \mathrm{eV}$ and $w_b=2\ \mathrm{eV}$. To compare the performance of the ACA to their results we show here for which truncation level the ACA is practically exact for this system. We have chosen this indirect method of comparison because in order to faithfully compare the two computational methods it would be necessary to know which truncation parameter in the ACA corresponds to their variational exact diagonalization method in terms of computational effort. 

\begin{figure}[htb]
 \includegraphics[width=\linewidth,height=!]{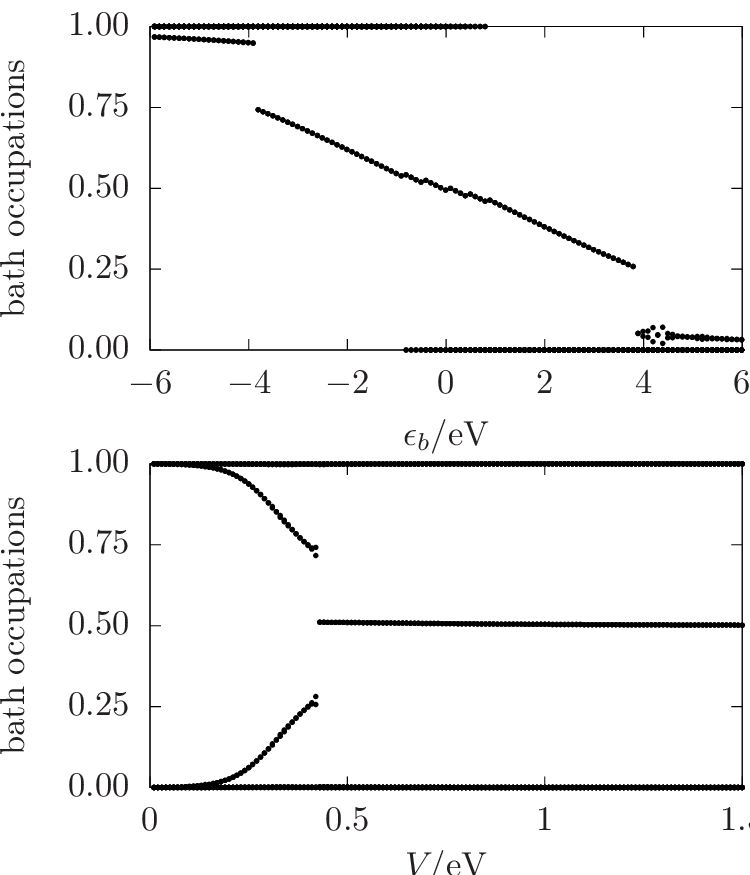}
  \caption{\label{fig:SIAMMALTEoccb} Eigenvalue spectrum (occupations) of the bath density matrix $\mat{\rho}_{\text{bath,bath,exact}}$ of the exact ground state. The model is the 7-site SIAM used by Sch\"uler et al.\cite{maltevared,Schueler2017}. The upper graph shows the eigenvalues for the model parameters used in~\cite{maltevared} and the lower one the corresponding ones for the parameter regime of~\cite{Schueler2017}. The model parameters are explained in section~\ref{sec:siam_beyond}. 
 }
\end{figure}

In their first work\cite{maltevared} Sch\"uler et al. fixed the hybridization strength $V_{1,j}=0.9\ \mathrm{eV}$ and varied the mean bath energy $\epsilon_b\in[-6\ \mathrm{eV},6\ \mathrm{eV}]$. For $\epsilon_b< -3.9 \ \mathrm{eV}$ and $\epsilon_b> 3.9  \ \mathrm{eV}$ we found a doubly degenerate ground state and otherwise a non-degenerate ground state.
In the subsequent work\cite{Schueler2017} they fixed the mean bath energy $\epsilon_b=0.02\ \mathrm{eV}$ and varied the hybridization strength $V=V_{1,j}\in [0.0\ \mathrm{eV},1.5\ \mathrm{eV}]$. For this choice the ground state is doubly degenerate below the hybridization strength $V\approx 0.42\ \mathrm{eV}$ and non-degenerate above. The eigenvalue spectra of the bath density matrix of the exact ground state for these two parameter ranges are shown in Fig.~\ref{fig:SIAMMALTEoccb}. For a non-degenerate ground state there are at most three clusters of eigenvalues and for a degenerate ground state at most four. These clusters have a finite but very small spread. Consequently the ACA with M=3 for the non-degenerate ground states or M=4 for the degenerate ground states would result in a ground state energy and one-particle density matrix with a negligible deviation from the exact results.

\subsection{\label{sec:siam_beyond_all}Approaching a continuous bath density of states }
For the model described by Eq.~\eqref{eq:SIAM2_H_imp}-Eq.~\eqref{eq:SIAM2_H_tot} we can easily increase the number of bath sites $L_\mathrm{bath}$. 
We investigate the interaction strength dependence of the spectrum of the bath density matrix of this model with $L_\mathrm{imp}=1$ for half-filling. The bandwidth of the bath was chosen similar to Sch\"uler et al. as $w_b=2\ \mathrm{eV}$ and $\epsilon_{f,1}=0$. The hybridization parameters $V_{1,j}$ are distributed randomly in the interval $[0,w_b/L_\mathrm{bath}]$ to break any accidental spatial symmetry of the model. 

\begin{figure}[htb]
 \includegraphics[width=\linewidth,height=!]{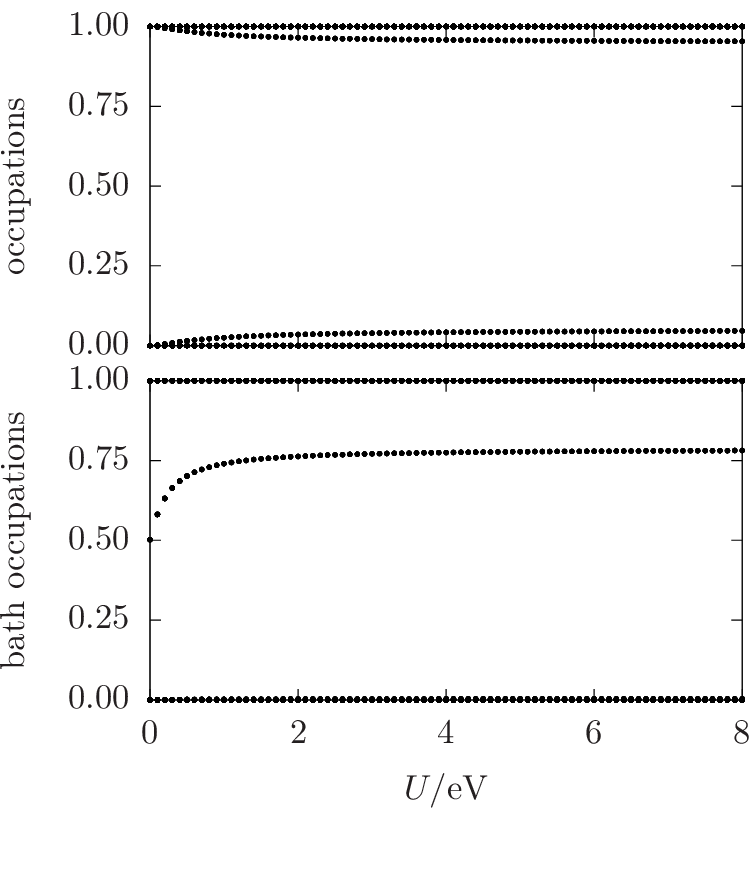}
  \caption{\label{fig:SIAMRandom1}  Eigenvalue spectrum (occupations) of the full one-particle reduced density matrix $\mat\rho_\mathrm{exact}$ (upper graph) and of the bath density matrix $\mat{\rho}_{\text{bath,bath,exact}}$ (lower graph) of the exact ground state of a SIAM with a single-site impurity ($L_\mathrm{imp}=1$, $L_\mathrm{imp}+L_\mathrm{bath}=200$). 
  All other parameters are described in section~\ref{sec:siam_beyond_all}.
 }
\end{figure}

For the SIAM with a single-site impurity ($L_\mathrm{imp}=1$) we used a total number of $L_\mathrm{imp}+L_\mathrm{bath}=200$ sites such that the spacing of the bath energy levels is approx. $0.01 \ \mathrm{eV}$. This is sufficiently close to the limit of a continuous density of states. The DMRG-results for the eigenvalues of the bath density matrix for several interaction parameters are shown in Fig.~\ref{fig:SIAMRandom1}. It is clearly visible that there are only three clusters of eigenvalues just like in the bath density matrix of the related model with only eleven bath sites shown in Fig.~\ref{fig:SIAMTLAUdepoccb}. Thus we expect this feature to hold also for a continuous density of states.

\begin{figure}[htb]
 \includegraphics[width=\linewidth,height=!]{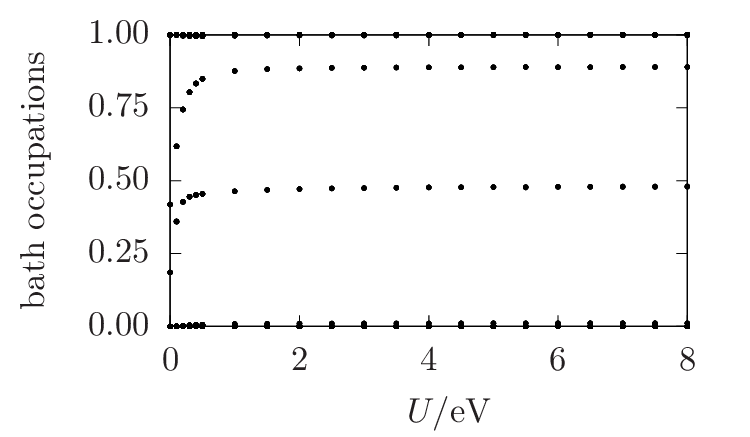}
  \caption{\label{fig:SIAMRandom2}  Eigenvalue spectrum (occupations) of the bath density matrix $\mat{\rho}_{\text{bath,bath,exact}}$ of the exact ground state of a SIAM with a two-site impurity ($L_\mathrm{imp}=2$, $L_\mathrm{imp}+L_\mathrm{bath}=40$).  
  All other parameters are described in section~\ref{sec:siam_beyond_all}.
 }
\end{figure}

\begin{figure}[htb]
 \includegraphics[width=\linewidth,height=!]{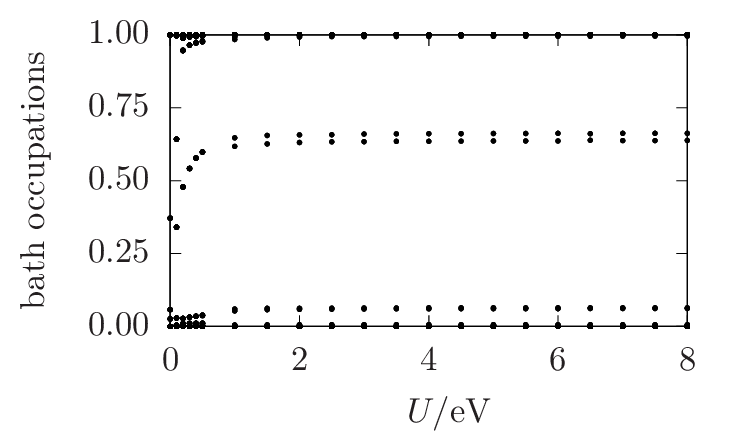}
  \caption{\label{fig:SIAMRandom3}  Eigenvalue spectrum (occupations) of the bath density matrix $\mat{\rho}_{\text{bath,bath,exact}}$ of the exact ground state of a SIAM with a three-site impurity ($L_\mathrm{imp}=3$, $L_\mathrm{imp}+L_\mathrm{bath}=40$).  
  All other parameters are described in section~\ref{sec:siam_beyond_all}. 
 }
\end{figure}

For the SIAM with a two-site-impurity ($L_\mathrm{imp}=2$) there are four distinct clusters of eigenvalues of the bath density matrix as shown in Fig.~\ref{fig:SIAMRandom2}. These results together with the five clusters emerging for an impurity with three sites shown in Fig.~\ref{fig:SIAMRandom3} indicate a linear dependency between the number of eigenvalue clusters and the impurity size. Thus the numerical results suggest for a SIAM with a non-degenerate ground state and a spin-independent one-particle Hamiltonian the relation 
\begin{equation}
  n_\mathrm{cluster}=2+L_\mathrm{imp}
\end{equation}
between the number of sites $L_\mathrm{imp}$ in the impurity and the number of eigenvalue cluster $n_\mathrm{cluster}$ in the bath density matrix. Cauchy's eigenvalue interlacing theorem\footnote{Cauchy's eigenvalue interlacing theorem is sometimes called Poincar\'e separation theorem.} states that the eigenvalues of a principal submatrix $B\in \mathbb{C}^{(n-1)\times (n-1)}$ of a hermitian matrix $A\in\mathbb{C}^{n\times n}$ and the eigenvalues of $A$ interlace. Applied to the situation here the theorem states that the eigenvalues $f_{\sigma,i}$ of the full one-particle reduced density matrix and the eigenvalues $f_{\mathrm{bath},\sigma,i}$ of the bath density matrix for a spin direction $\sigma\in\{\uparrow,\downarrow\}$ fulfill the relation
\begin{equation}
f_{\sigma,1} \leq f_{\mathrm{bath},\sigma,1} \leq f_{\sigma,2} \leq .... \leq f_{\sigma,n-1} \leq f_{\mathrm{bath},\sigma,n-1} \leq f_{\sigma,n}.
\end{equation}
As most eigenvalues of the full one-particle reduced density matrix of a SIAM are zero or one this relation requires that also most eigenvalues of the bath density matrix are zero or one. However according to the interlacing theorem there could be up to five distinct clusters of eigenvalues of the bath density matrix. This is the case because the eigenvalues of the full density matrix are arranged in four clusters as shown in Fig.~\ref{fig:SIAMRandom1} where the clusters with fractional eigenvalues only have a two-fold degeneracy because of the two spin directions. Why the number of eigenvalue clusters is smaller than the number allowed by Cauchy's eigenvalue interlacing theorem is a interesting topic for further research. 

In conclusion, the ACA would be practically exact for this model for the truncation parameter $M=n_\mathrm{cluster}=2+L_\mathrm{imp}$. The clusters with fractional occupations are only two-fold degenerate, $n_j=2$, so that the effective bath level corresponding to these eigenvalues contains only two instead of $N_\mathrm{imp}=2L_\mathrm{imp}$ spin-orbitals as discussed in section~\ref{sec:exact_limits}. Consequently for a truncation parameter of $M=2+L_\mathrm{imp}$ the ACA only constructs $2(n_\mathrm{cluster}-2)+2N_\mathrm{imp}=3N_\mathrm{imp}\leq MN_\mathrm{imp}$ effective bath states. 

\section{\label{sec:conclusion}Conclusion}
We  have   introduced a method, which we call adaptive cluster approximation (ACA), to evaluate the density-matrix functional for a single impurity Anderson model. 
The ACA provides a systematic way to reduce the number of bath sites surrounding the impurity with minimal loss of accuracy.
For this smaller cluster, the density-matrix functional can be evaluated using Levy's constrained-search algorithm or other advanced approaches with an unfavourable scaling of the computational complexity. This is an important step towards the use of advanced density-matrix functionals in first-principles calculations, especially in the context of hybrid theories combining density functional theory and local rDMFT\cite{bloechl11_prb84_205101,epjstrdmft}.

An effective correction scheme has been presented, which reduces the build-up of truncation errors during optimization of the density matrix. These deviations result from the absence of constraining forces on the corresponding density matrix elements. This correction scheme uses parameterized functionals to embed the truncated cluster into a larger effective system

For the lowest possible truncation, that is one effective bath site, and a single interacting site the ACA is equivalent to the two-level approximation of T\"ows et al.\cite{toews11_prb83_235101}. In contrast to the two-level approximation, the ACA can be applied to multi-orbital impurities and systematically converged to the exact result. The performance of the ACA has been explored for a single-orbital SIAM with a finite bath. The results show that the ground-state energy and orbital occupations converge rapidly with the level of the effective bath in the ACA.

\begin{acknowledgements}
   We would like to thank Thomas K\"ohler and Sebastian Paeckel for fruitful discussions about MPS-DMRG calculations.  
Financial support from the Deutsche Forschungsgemeinschaft
through FOR1346 (project 9) is gratefully acknowledged.
\end{acknowledgements}

\appendix
\section{\label{appendix:proof_truncation}}
Here we show that the density-matrix functional of a block-diagonal one-particle reduced density matrix is the sum of the density-matrix functional of the interacting block and an entropy contribution from the non-interacting block. That is, for an interaction $\hat W$ restricted to the impurity states $\mathrm{imp}$, the relation
\begin{equation}
\label{eq:appendix_rho}
  F_\beta^{\hat W}\left [\begin{pmatrix} \mat{\rho}_{\mathrm{imp}}  & 0 \\ 0 & \mat{\rho}_{\mathrm{bath}}\end{pmatrix}\right]=F_\beta^{\hat W}\left [{\mat{\rho}_\mathrm{imp}}\right]+F^{\hat 0}_{\beta}[\mat{\rho}_\mathrm{bath}],
\end{equation} 
holds. For the purpose of this proof, the impurity can also include some non-interacting orbitals and we work at finite temperature to avoid some non-uniqueness problems\cite{gilbert75_prb12_2111,baldsiefen15_pra92_052514,doi10.10631.4927075,2017arXiv171008805G}. The inverse temperature is defined as $\beta=1/(k_BT)$.

First we note some useful properties of the grand potential 
\begin{equation}
  \label{eq:appendix_omega_def}
\Omega_{\beta,\mu}^{\hat W}[\mat{h}]=-\frac{1}{\beta} \ln \left(\mathrm{Tr} e^{-\beta(\hat h+\hat W-\mu \hat N)}\right),
\end{equation}
where $\mu$ denotes the chemical potential and $\hat N$ the operator of the total particle number.
The grand potential is a concave function\cite{PhysRevB.82.205120,2017arXiv171008805G} of the matrix elements $\mat{h}$ of the one-particle Hamiltonian $\hat{h}$, that is 
\begin{equation}
\Omega_{\beta,\mu}^{\hat W}[(1-\lambda)\mat{h}_1+\lambda \mat{h}_2]\geq (1-\lambda) \Omega_{\beta,\mu}^{\hat W}[\mat{h}_1]+\lambda \Omega_{\beta,\mu}^{\hat W}[\mat{h}_2].
\end{equation}
At finite temperatures the above inequality is strict, i.e. the grand potential is strictly concave.
Additionally the grand potential is an extensive quantity,
\begin{equation}
  \label{appendix:omega_ext}
  \Omega_{\beta,\mu}^{\hat W}\left [\begin{pmatrix} \mat{h}_{\mathrm{imp}}  & 0 \\ 0 & \mat{h}_{\mathrm{bath}}\end{pmatrix}\right]=\Omega_{\beta,\mu}^{\hat W}\left [ \mat{h}_{\mathrm{imp}} \right]+\Omega_{\beta,\mu}^{\hat 0}\left [ \mat{h}_{\mathrm{bath}} \right].
\end{equation}
This can easily be shown by using that $\hat{h}_{\mathrm{imp}}+\hat W$ and $\hat{h}_{\mathrm{bath}}$ commute.

The derivatives of the grand potential with respect to the matrix elements $\mat h$ of the one-particle Hamiltonian $\hat h$ are given by the thermal expectation values of the one-particle density-matrix operator as
\begin{equation}
  \label{eq:appendix_omega_deriv}
  \frac{\partial \Omega^{\hat W}_{\beta,\mu} \left[\mat h\right]}{\partial h_{i,j}} =\frac{\mathrm{Tr}\left( e^{-\beta(\hat h+\hat W-\mu \hat N)} \hat c_i^\dagger \hat c_j\right)}{\mathrm{Tr} e^{-\beta(\hat h+\hat W-\mu \hat N)}}. 
\end{equation}

We can apply the general relation Eq.~\eqref{eq:appendix_omega_deriv} to the situation of a one-particle Hamiltonian with block-diagonal matrix elements
\begin{equation}
 \mat h=\begin{pmatrix}\mat{h}_{\mathrm{imp}}  & 0 \\ 0 & \mat{h}_{\mathrm{bath}}\end{pmatrix}.
\end{equation}
Then the Hamiltonian $\hat h_{\mathrm{imp}}+\hat W$, which only acts on the impurity states, commutates with the one-particle Hamiltonian $\hat h_{\mathrm{bath}}$ of the bath.
The eigenstates $|\Psi_{a,b}\rangle$ of the full system,
\begin{eqnarray}
  (\hat h_{\mathrm{imp}}+\hat W+\hat h_{\mathrm{bath}})|\Psi_{a,b}\rangle= \epsilon_{a,b} |\Psi_{a,b}\rangle,\\
  \hat N |\Psi_{a,b}\rangle= n_{a,b} |\Psi_{a,b}\rangle,
\end{eqnarray}
can be written as product states of the eigenstates of the impurity Hamiltonian and eigenstates of the bath Hamiltonian,
\begin{eqnarray}
  (\hat h_{\mathrm{imp}}+\hat W)|\Psi_{\mathrm{imp},a}\rangle&=& \epsilon_{\mathrm{imp},a} |\Psi_{\mathrm{imp},a}\rangle\\
  \hat N|\Psi_{\mathrm{imp},a}\rangle&=& n_{\mathrm{imp},a} |\Psi_{\mathrm{imp},a}\rangle,\\
  \hat h_{\mathrm{bath}}|\Psi_{\mathrm{bath},b}\rangle&=& \epsilon_{\mathrm{bath},b} |\Psi_{\mathrm{bath},b}\rangle,\\
  \hat N|\Psi_{\mathrm{bath},b}\rangle&=& n_{\mathrm{bath},b} |\Psi_{\mathrm{bath},b}\rangle,
\end{eqnarray}
as $|\Psi_{a,b}\rangle=|\Psi_{\mathrm{imp},a}\rangle \otimes |\Psi_{\mathrm{bath},b}\rangle$.
Expectation values of the one-particle reduced density-matrix operator between impurity and bath states vanish for such product states,
\begin{eqnarray}
  \nonumber
  \langle \Psi_{a,b}|\hat c^\dagger_\mathrm{imp} \hat c_\mathrm{bath} | \Psi_{a,b}\rangle&=&\langle\Psi_{\mathrm{imp},a}| \hat c^\dagger_\mathrm{imp}|\Psi_{\mathrm{imp},a}\rangle\\
  \nonumber  &\cdot&  \langle\Psi_{\mathrm{bath},b}|\hat c_\mathrm{bath}|\Psi_{\mathrm{bath},b}\rangle\\
  &=&0.
\end{eqnarray}
The matrix elements vanish because the eigenstates $|\Psi_{\mathrm{imp},a}\rangle$ and $|\Psi_{\mathrm{bath},b}\rangle$ are also eigenstates of the total particle number operator. Consequently also the expectation values 
\begin{widetext}
\begin{equation}
  \frac{\mathrm{Tr}\left( e^{-\beta(\hat h+\hat W-\mu \hat N)} \hat c^\dagger_\mathrm{imp} \hat c_\mathrm{bath}\right)}{\mathrm{Tr} e^{-\beta(\hat h+\hat W-\mu \hat N)}} =\sum_{a,b} \frac{e^{-\beta(\epsilon_{a,b}-\mu n_{a,b})}}{\sum_{{a',b'}} e^{-\beta(\epsilon_{a',b'}-\mu n_{a',b'})}} \langle \Psi_{a,b}|\hat c^\dagger_\mathrm{imp} \hat c_\mathrm{bath} | \Psi_{a,b}\rangle
\end{equation}
\end{widetext}
  vanish in the thermal ground state ensemble. Thus, for a vanishing impurity-bath Hamiltonian
 the derivatives of the grand potential with respect to the matrix elements of the impurity-bath Hamiltonian vanish, i.e.
\begin{equation}
  \frac{\partial \Omega^{\hat W}_{\beta,\mu} \left[\begin{pmatrix}\mat{h}_{\mathrm{imp}}  & \mat{h}_{\mathrm{imp,bath}} \\ \mat{h}_{\mathrm{imp,bath}}^\dagger & \mat{h}_{\mathrm{bath}}\end{pmatrix}\right]}{\partial (\mat h_{\mathrm{imp,bath}})_{i,j}}\Bigg |_{\mat h_{\mathrm{imp,bath}}=\mat 0}=0 \ \forall i,j.
\end{equation}
This means that $\mat{h}_\mathrm{imp,bath}=0$ is a stationary point of the maximization problem 
\begin{equation}
  \label{appendix:omega_max}
  \max_{\mat{h}_\mathrm{imp,bath}}\Omega_{\beta,\mu}^{\hat W}\left[\begin{pmatrix} \mat{h}_{\mathrm{imp}}  & \mat{h}_{\mathrm{imp,bath}} \\ \mat{h}_{\mathrm{imp,bath}}^\dagger & \mat{h}_{\mathrm{bath}}\end{pmatrix} \right].
\end{equation}
Combined with the concavity of the grand potential with respect to the matrix elements of the one-particle Hamiltonian this shows that $\mat{h}_\mathrm{imp,bath}=0$ is a maximizer of the problem in Eq.~\eqref{appendix:omega_max}.

For the main part of the proof, that is to show Eq.~\eqref{eq:appendix_rho}, we use that the density-matrix functional is the Legendre-Fenchel transform of the grand potential\cite{lieb83_ijqc24_243} $\Omega_{\beta,\mu}^{\hat W}(\mat{h})$ with respect to the one-particle Hamiltonian
\begin{eqnarray}
  \label{eq:appendix_Frho}
  F^{\hat W}_{\beta}[\mat{\rho}]&=&\max_{\mat{h},\mu} \left\{\Omega_{\beta,\mu}^{\hat W}[\mat{h}]- \mathrm{Tr}(\mat{\rho}\mat{h}) \right\}.
\end{eqnarray}
For the block-diagonal density matrix in Eq.~\eqref{eq:appendix_rho} the density-matrix functional is
\begin{widetext}
\begin{equation}
  \label{eq:appendix_Frho2}
  F^{\hat W}_{\beta}\left [\begin{pmatrix} \mat{\rho}_{\mathrm{imp}}  & 0 \\ 0 & \mat{\rho}_{\mathrm{bath}}\end{pmatrix}\right]=\max_{\mat{h}_\mathrm{imp},\mat{h}_\mathrm{bath},\mat{h}_\mathrm{imp,bath},\mu} \left\{\Omega_{\beta,\mu}^{\hat W}\left[\begin{pmatrix} \mat{h}_{\mathrm{imp}}  & \mat{h}_{\mathrm{imp,bath}} \\ \mat{h}_{\mathrm{imp,bath}}^\dagger & \mat{h}_{\mathrm{bath}}\end{pmatrix}\right]- \mathrm{Tr}(\mat{\rho}_\mathrm{imp}\mat{h}_\mathrm{imp})-\mathrm{Tr}(\mat{\rho}_\mathrm{bath}\mat{h}_\mathrm{bath}) \right\}.
\end{equation}
\end{widetext}
The maximization over $\mat{h}_\mathrm{imp,bath}$ can be performed with the help of Eq.~\eqref{appendix:omega_max}, which yields together with Eq.~\eqref{appendix:omega_ext} the decomposition
\begin{widetext}
\begin{eqnarray}
  \label{eq:appendix_Frhosplit}
  F^{\hat W}_{\beta}\left [\begin{pmatrix} \mat{\rho}_{\mathrm{imp}}  & 0 \\ 0 & \mat{\rho}_{\mathrm{bath}}\end{pmatrix}\right]&=&\max_{\mat{h}_\mathrm{imp},\mat{h}_\mathrm{bath},\mu} \Bigg\{\Omega_{\beta,\mu}^{\hat W}\left[\begin{pmatrix} \mat{h}_{\mathrm{imp}}  &0\\0 & \mat{h}_{\mathrm{bath}}\end{pmatrix}\right]- \mathrm{Tr}(\mat{\rho}_\mathrm{imp}\mat{h}_\mathrm{imp})-\mathrm{Tr}(\mat{\rho}_\mathrm{bath}\mat{h}_\mathrm{bath}) \Bigg\}\\
  &=&\max_{\mat{h}_\mathrm{imp},\mu} \Bigg\{\Omega_{\beta,\mu}^{\hat W}[\mat{h}_\mathrm{imp}]- \mathrm{Tr}(\mat{\rho}_\mathrm{imp}\mat{h}_\mathrm{imp}) \Bigg\}+\max_{\mat{h}_\mathrm{bath},\mu} \Bigg\{\Omega_{\beta,\mu}^{\hat 0}[\mat{h}_\mathrm{bath}]- \mathrm{Tr}(\mat{\rho}_\mathrm{bath}\mat{h}_\mathrm{bath}) \Bigg\}\\
  &=& F^{\hat W}_{\beta}[\mat{\rho}_\mathrm{imp}]+F^{\hat 0}_{\beta}[\mat{\rho}_\mathrm{bath}].
\end{eqnarray}
\end{widetext}
This concludes the proof. The non-interacting density-matrix functional $F^{\hat 0}_{\beta}[\mat{\rho}_\mathrm{bath}]$ only contains an entropy contribution
\begin{equation}
  F^{\hat 0}_{\beta}[\mat{\rho}_\mathrm{bath}]=\frac{1}{\beta} \mathrm{Tr} \left[ \mat{\rho}_\mathrm{bath} \mathrm{ln}(\mat{\rho}_\mathrm{bath})+(1-\mat{\rho}_\mathrm{bath}) \mathrm{ln}(1-\mat{\rho}_\mathrm{bath}) \right]
\end{equation}
and vanishes in the zero temperature limit.

%

\end{document}